%% file: main.tex
\documentclass[conference]{IEEEtran}
\ifCLASSINFOpdf
\else
\fi
\usepackage[table]{xcolor}  
\usepackage{amsmath}
\usepackage{graphicx}
\usepackage{booktabs}
\usepackage{latexsym} 
\usepackage{array}    
\usepackage{multirow}
\usepackage{subcaption}
\usepackage{algorithm}
\usepackage[noend]{algpseudocode}
\usepackage{threeparttable}
\usepackage{paralist}   
\usepackage{xspace}
\usepackage{color}
\usepackage{adjustbox}
\usepackage{balance}
\usepackage{wasysym}
\usepackage{rotating}   
\usepackage{listings}
\usepackage{tabu}
\usepackage{pifont}
\usepackage{framed}
\usepackage{url}            
\usepackage{hyperref}

\usepackage{cite}

\usepackage{xpatch}
\makeatletter
\chardef\TPT@@@asteriskcatcode=\catcode`*
\catcode`*=11
\xpatchcmd{\threeparttable}
  {\TPT@hookin{tabular}}
  {\TPT@hookin{tabular}\TPT@hookin{tabu}}
  {}{}
\catcode`*=\TPT@@@asteriskcatcode
\makeatother

\definecolor{wheat1}{rgb}{1.000000,0.905882,0.729412}
\definecolor{LightGray}{rgb}{0.827451,0.827451,0.827451}

\newcolumntype{a}{>{\columncolor{wheat1}}l}

\definecolor{mygreen}{rgb}{0,0.6,0}
\definecolor{mygray}{rgb}{0.5,0.5,0.5}
\definecolor{mymauve}{rgb}{0.58,0,0.82}
\definecolor{darkblue}{rgb}{0.0,0.0,0.6}
\definecolor{maroon}{RGB}{102, 0, 0}
\definecolor{Maroon}{cmyk}{0,0.87,0.68,0.32}
\definecolor{darkred}{RGB}{139, 0, 0}
\definecolor{forestgreen}{RGB}{34, 139, 34}

\lstdefinelanguage{XML}
{
  basicstyle=\ttfamily\small,   
  morestring=[b]",
  moredelim=[s][\color{darkblue}]{<}{\ },
  moredelim=[s][\color{darkblue}]{</}{>},
  moredelim=[l][\color{darkblue}]{/>},
  moredelim=[l][\color{darkblue}]{>},
  morecomment=[s]{<?}{?>},
  morecomment=[s]{<!--}{-->},
  stringstyle=\color{darkred},
  identifierstyle=\color{mymauve}
}

\lstdefinestyle{customJava}{
  breaklines=true,
  keepspaces=true,
  frame=single,
  language=Java,
  showstringspaces=false,
  basicstyle=\footnotesize\ttfamily,
  keywordstyle=\color{blue},
  otherkeywords={+, getIntent},
  numbers=left,
  numbersep=5pt,
  numberstyle=\scriptsize\color{black},
  rulecolor=\color{black},
  stepnumber=1,
  tabsize=2,
  commentstyle=\itshape\color{green!40!black},
  stringstyle=\color{orange},
  emph=[1]  
  {
        do,
        try,
        new,
        catch,
        while,
        SecProvider,
        SecReceiver,
        SecService,
        SecActivity,
        SecSink,
  },
  emphstyle=[1]{\color{darkred}},
  emph=[2]  
  {
        @Override,
  },
  emphstyle=[2]{\color{purple!40!black}},
  belowskip=-1em, 
}

\newif\ifANNOYMIZE
\ANNOYMIZEfalse

\newif\ifACM
\ACMfalse 

\ifACM
\fi

\ifACM
\newcommand{\myfig}{Figure\xspace}
\else
\newcommand{\myfig}{Fig.\xspace}
\fi

\newcommand{\mytab}{Table\xspace}

\ifACM
\newcommand{\mysec}{\S}
\else
\newcommand{\mysec}{Sec.\xspace}
\fi

\newcommand{\name}{BlockScope\xspace}
\newcommand{\eth}{Ethereum\xspace}
\newsavebox{\bigimage} 

\begin{document}
%
\title{\name: Investigating Propagated Vulnerabilities and Their Patching Processes in Forked Blockchain Projects}
\title{\name: Detecting and Investigating Propagated Vulnerabilities in Forked Blockchain Projects}



%

\author{
\IEEEauthorblockN{Xiao Yi$^1$, Yuzhou Fang$^1$, Daoyuan Wu$^1$$^*$\thanks{$^*$Corresponding author.}, and Lingxiao Jiang$^2$}
\IEEEauthorblockA{$^1$The Chinese University of Hong Kong\\
$^2$Singapore Management University}
}


\IEEEoverridecommandlockouts
\makeatletter\def\@IEEEpubidpullup{6.5\baselineskip}\makeatother
\IEEEpubid{\parbox{\columnwidth}{
    Network and Distributed System Security (NDSS) Symposium 2023\\
    27 February - 3 March 2023, San Diego, CA, USA\\
    ISBN 1-891562-83-5\\
    https://dx.doi.org/10.14722/ndss.2023.24222\\
    www.ndss-symposium.org
}
\hspace{\columnsep}\makebox[\columnwidth]{}}

\maketitle

\input{abstract}

\input{intro}
\input{backg}
\input{method}

\input{detect}

\input{investigate}

\input{discuss}

\input{related}
\input{conclude}
\section*{Acknowledgment}
We would like to thank all the reviewers for their valuable comments and constructive suggestions to this paper.
This work was partially supported by a direct grant (ref. no. 4055127) from The Chinese University of Hong Kong.

\balance
\bibliographystyle{IEEEtranS}   
\bibliography{main}

\input{appendix}

\end{document}

%% file: abstract.tex
\begin{abstract}

Due to the open-source nature of the blockchain ecosystem, it is common for new blockchains to fork or partially reuse the code of classic blockchains.
For example, the popular Dogecoin, Litecoin, Binance BSC, and Polygon are all variants of Bitcoin/\eth. 
These ``forked'' blockchains thus could encounter similar vulnerabilities that are propagated from Bitcoin/Ethereum during forking or subsequently commit fetching.
In this paper, we conduct a systematic study of detecting and investigating the propagated vulnerabilities in forked blockchain projects.
To facilitate this study, we propose \name, a novel tool that can effectively and efficiently detect multiple types of cloned vulnerabilities given an input of existing Bitcoin/\eth security patches.
Specifically, \name adopts similarity-based code match and designs a new way of calculating code similarity to cover all the syntax-wide variant (i.e., Type-1, Type-2, and Type-3) clones.
Moreover, \name automatically extracts and leverages the contexts of patch code to narrow down the search scope and locate only potentially relevant code for comparison. 

Our evaluation shows that \name achieves good precision and high recall both at 91.8\% (1.8 times higher recall than that in the state-of-the-art ReDeBug while with close precision).
\name allows us to discover 101 previously unknown vulnerabilities in 13 out of the 16 forked projects of Bitcoin and \eth, including 16 from Dogecoin, 6 from Litecoin, 1 from Binance BSC, and 4 from Optimism. 
We have reported all the vulnerabilities to their developers; 40 of them have been patched or accepted, 66 were acknowledged or under pending, and only 4 were rejected.
We further investigate the propagation and patching processes of discovered vulnerabilities, and reveal three types of vulnerability propagation from source to forked projects, as well as the long delay (mostly over 200 days) for releasing patches in Bitcoin forks (vs. $\sim$100 days for Ethereum forks).

\end{abstract}

%% file: intro.tex

\section{Introduction}
\label{sec:intro}

Blockchain~\cite{saad2019exploring} and DeFi (Decentralized Finance)~\cite{werner2021sok} are emerging in recent years.
A good development in the blockchain ecosystem is that many projects are open-source.
This is particularly true for the public blockchains like Bitcoin and \eth.
As a result, new blockchains could fork or partially reuse the code of classic blockchains to speed up the development.
Notably, Bitcoin is the one with most forked projects --- the popular Dogecoin, Litecoin, Dash, Zcash, and Bitcoin Cash/SV are all variants of Bitcoin.
In recent years, \eth was also forked by a number of EVM (\eth Virtual Machine)-compatible chains, such as Binance Smart Chain (BSC), Polygon, Avalanche Contract Chain, and Optimism (\eth's Layer-2 rollup network).

However, ``forked'' blockchains could encounter similar vulnerabilities that appeared in the code of Bitcoin and \eth.
Specifically, a vulnerability could be propagated from Bitcoin/\eth to the forked projects during the initial fork or subsequently when updated commits are fetched from Bitcoin/\eth.
In this paper, we aim to systematically detect cloned vulnerabilities in forked blockchain projects and investigate how they are propagated and patched.

To facilitate this study and future analysis, we propose \name, a novel tool that can not only automatically detect vulnerable clones but also pinpoint the cases already fixed and their patching process information.
To achieve effective and efficient detection on all the syntax-wide cloned vulnerabilities (i.e., Type-1, Type-2, and Type-3 clones, as to be defined in \mysec\ref{subsec:clone_types}), \name has two unique designs as compared to typical code clone detection tools, e.g.,~\cite{Redebug2012, VUDDY2017, MVP2020, CCFinder02, CPMiner04, Sourcercc2016}. 
First, we adopt similarity-based code match, instead of the hash-based exact match in ReDeBug~\cite{Redebug2012}, VUDDY~\cite{VUDDY2017}, and MVP~\cite{MVP2020}, so that \name is more tolerant to the code lines with no exact ``abstracted'' hashes.
Moreover, we design a new way of calculating code similarity to better handle the code fragments with inserted/deleted/reordered code lines. 
According to our evaluation with the state-of-the-art ReDeBug tool, our new design greatly reduces false negatives while only slightly increasing false positives for our problem. 
Second,
\name automatically extracts and leverages patch code contexts to locate only potentially relevant code for comparison. 
This not only dramatically improves the running performance for large projects, e.g., 15.4 times faster than ReDeBug in analyzing \eth's forked projects with more lines of code (LOC), but also enhances the detection precision because the context similarity is also being considered.

To evaluate \name, we collect a dataset of 38 security patches --- 32 of them are directly from Bitcoin's repository because there were only four CVEs in the recent five years, and the rest six are CVEs of Ethereum reported in the last three years.
With this input, we apply \name and ReDebug to test 11 most popular forked projects of Bitcoin and 5 of Ethereum (identified from nearly the top 100 cryptocurrencies), with 4.2M C/C++ LOC and 3.5M Go LOC, respectively. 
The evaluation shows that BlockScope detects 101 true vulnerabilities in all the 13 forked projects (three projects, Qtum, Avalanche, and Polygon, does not contain any of the tested vulnerabilities), whereas ReDeBug detects only 57 vulnerabilities in 11 forked projects.
By performing a thorough code review of all the raw detection results, we find that \name achieves good precision and high recall both at 91.8\%, whereas ReDeBug's recall is only 51.8\% despite its precision at 95\%. 
Among the 101 vulnerabilities automatically detected by \name, we are able to identify serious ones from the top blockchains like Dogecoin, Litecoin, Bitcoin SV, Binance BSC, and Optimism.
This demonstrates the real-world impact of our work\footnote{Binance acknowledged our vulnerability report with a bug bounty reward.}.

We further investigate how the discovered vulnerabilities\footnote{Besides 101 automatically detected cases, we also analyzed 9 that were false negatives in \name but manually identified during the evaluation.} are propagated from Bitcoin/\eth to their forked projects and understand the patching processes of the 138 cases that were already fixed in forked projects before our detection.
Specifically, we reveal three types of vulnerability propagation from Bitcoin/\eth to their forked projects, including the cases directly forked in the beginning, fetched from vulnerable commits, and infected with no explicitly vulnerable commits.
Besides vulnerability propagation, we additionally identify three other propagation that caused false positives and negatives in \name; details in \mysec\ref{subsec:false_detection_analysis}. 
As for patch delays, we find that only DigiByte, among the six forked projects of Bitcoin with enough patched cases, can catch up with Bitcoin's patch release schedule.
The patch delays for the other five are typically long, mostly over 200 days.
Compared with Bitcoin, the result for \eth's forked projects is relatively acceptable, with half of the patches released within 100 days.

\textbf{Contributions.}
To sum up, we make the following major contributions in this paper:
\begin{itemize}
    \item \textit{(Methodology)} We propose novel patch-based clone detection for vulnerable code clones in forked projects, in which we design (i) a context-based search with similarity measurement to efficiently locate candidate code clones and (ii) a new way of calculating the similarity between two code fragments that is immune to Type-1/2/3 clones.

    \item \textit{(Detection)} We apply this methodology to detect 101 previously unknown vulnerabilities in the forked projects of Bitcoin and \eth with high precision and recall.

    \item \textit{(Investigation)} We further conduct a deep investigation of the vulnerability propagation and patching processes of the discovered vulnerabilities, and reveal new findings. 
\end{itemize}

\textbf{Ethics.}
As an ethical research and one contribution of this paper, we have spent significant efforts reporting all the 110 vulnerabilities (including nine false negatives manually identified during the evaluation).
The details are available in \mysec\ref{sec:report} and this GitHub repository, \url{https://github.com/VPRLab/BlkVulnReport}.

\textbf{Roadmap.}
The rest of this paper is organized as follows.
After explaining different blockchain projects and code clone types in \mysec\ref{sec:backg}, we first propose the \name tool in \mysec\ref{sec:method} to effectively detect the propagated vulnerabilities in the forked blockchains.
We then evaluate the accuracy and performance of \name and leverage it to discover previously unknown vulnerabilities in \mysec\ref{sec:detect}.
We further analyze how the discovered vulnerabilities are propagated from Bitcoin and \eth to the forked projects and understand their patching processes in \mysec\ref{sec:investigate}.
We then discuss some insights and implications in \mysec\ref{sec:discuss}.
Lastly, \mysec\ref{sec:related} reviews the related work and \mysec\ref{sec:conclude} concludes the paper.

%% file: backg.tex
\vspace{-2ex}
\section{Background}
\label{sec:backg}
\vspace{-1ex}

In this section, we first introduce the background of Bitcoin, Ethereum, and their popular forked projects in \mysec\ref{subsec:bitcoin_forks} and \mysec\ref{subsec:ethereum_forks}, and then provide the definition of different code clone types in \mysec\ref{subsec:clone_types}.

\subsection{Bitcoin and its Forked Projects}
\label{subsec:bitcoin_forks}

\textbf{Bitcoin} (BTC) \cite{BitcoinWhitePaper} is the first cryptocurrency that introduced the blockchain technology to the world.
Bitcoin leverages blockchain as a distributed ledger to guarantee the consensus between different peers.
Currently, Bitcoin is, without doubt, the dominant cryptocurrency, whose market capitalization takes around 40\% of the whole market.
Since Bitcoin is open-sourced, it has nourished many blockchain projects.
Specifically, among the top 100 cryptocurrencies on CoinMarketCap~\cite{CoinCap} as of 7 September 2021, we identified that 11 projects directly fork or partially reuse the code of Bitcoin.
We list them in \mytab~\ref{tab:repo_info_bit} and refer to them as Bitcoin's forked projects in this paper.

\begin{figure}[t!]
	\vspace{1ex}
	\begin{adjustbox}{center}
		\includegraphics[width=1.0\linewidth]{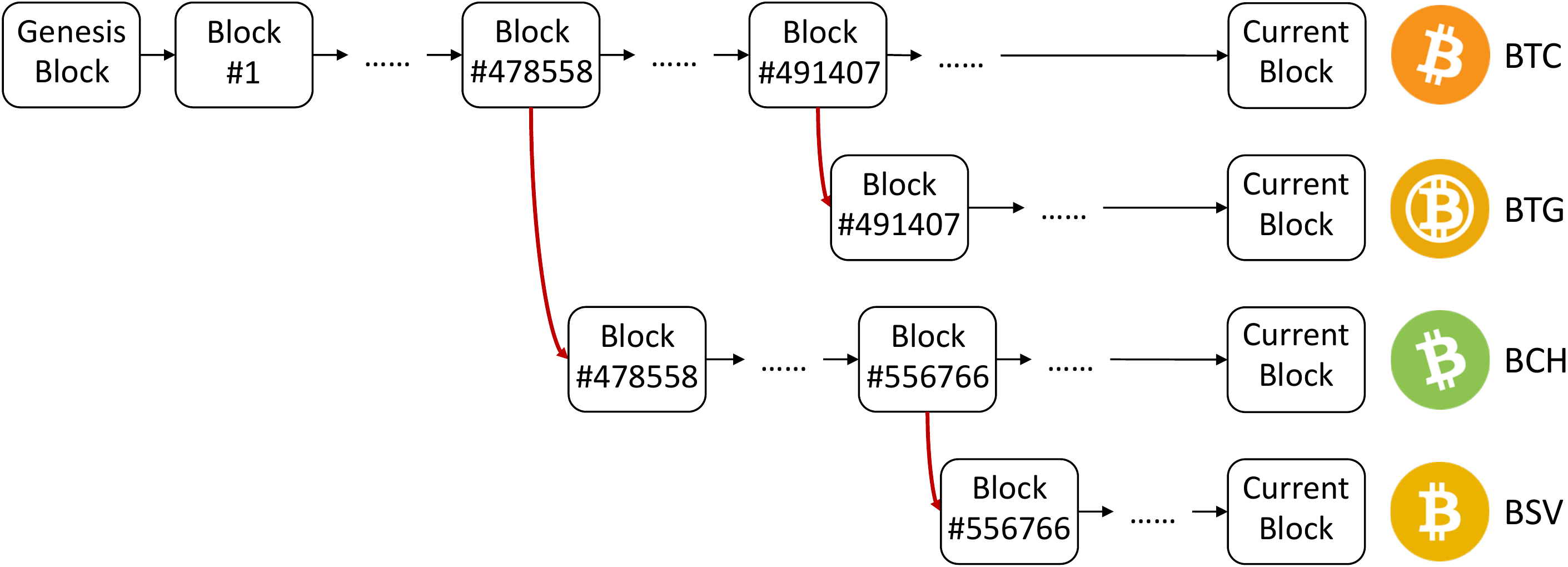}
	\end{adjustbox}
	\caption{Representative hard forks of Bitcoin.}
	\vspace{-1ex}
	\label{fig:bitcoin_forks}
\end{figure}

\input{resources/tab/tab_repo_info}

Most forked projects forked only the Bitcoin code, whereas \textbf{Bitcoin Cash} (BCH), \textbf{Bitcoin SV} (BSV), and \textbf{Bitcoin Gold} (BTG) also forked Bitcoin's blockchain, i.e., copying its transaction history, as the basis for their own blockchain~\cite{BTC_FORK}.
They are known as the ``hard forks'' of Bitcoin, as each of them creates a permanent fork of the original Bitcoin's blockchain.
We present the relationship between Bitcoin and these three projects in \myfig~\ref{fig:bitcoin_forks}.
As we can see, Bitcoin Cash is the earliest fork, which aims to reduce the transaction fee and improve the transaction speed of the original Bitcoin.
Therefore, they extend the maximum block size to 32MB, while the original Bitcoin's block size limit is 1MB.
Bitcoin SV further extends this limit to 128MB, which leads to another hard fork.
Bitcoin Gold, on the other hand, claims to solve the original Bitcoin's monopolized mining problem.
Specifically, they hope that by enabling mining on commonly available GPUs instead of specialized ASICs, it can democratize and decentralize the mining.

\textbf{Litecoin} gets its name from ``the light version of Bitcoin''.
Its goal is to provide faster transactions than Bitcoin.
Notably, instead of using Bitcoin's SHA-256, Litecoin adopts Scrypt~\cite{Scrypt} as the hash function, which offers a less compute-intensive but more memory-intensive mining process~\cite{LTCVSBTC}.
\textbf{Dogecoin} also leverages Scrypt as the hash function.
Indeed, it copies both Bitcoin's and Litecoin's code.
Although Dogecoin reached a market capitalization of over 40 billion USD, it was initially created as a meme cryptocurrency with an unlimited total supply~\cite{DOGEVSBTC}. 
\textbf{DigiByte} is another fork of Litecoin's code.
Besides SHA-256 and Scrypt, it can work with three more mining algorithms~\cite{DIGIEXP}.

\textbf{Dash} is not only a cryptocurrency but also a decentralized autonomous organization run by a subset of its users called ``masternodes''.
Specifically, anyone with 1,000 Dash can become a masternode in the Dash network and share the block reward.
Besides the standard node functions, the masternodes can vote on proposals to improve the ecosystem and provide two additional kinds of transactions, i.e., ``InstantSend'' and ``PrivateSend'' for instant transactions and private transactions, respectively~\cite{DASHVSBTC}.

\textbf{Zcash} and \textbf{Horizen} are designed to enhance the privacy for their users.
As the original Bitcoin is pseudo-anonymous, it is possible to decipher the patterns and connections involved, which may expose all information related to the sender and the receiver~\cite{ZECVSBTC}.
To tackle this problem, Zcash applies Zero-Knowledge proof algorithms (called zk-SNARKs) to ``shield'' the transactions so that it will not disclose the information about the coin holders.
Similarly, Horizen (formerly known as ZenCash) is a derivative of Zcash.
On top of the zk-SNARKs system, Horizen adopts a different funding model, which shares the block reward among miners, developers, and secure/super node operators, while Zcash just rewards miners and developers~\cite{ZECVSBTC}.

\textbf{Qtum} is a hybrid blockchain that combines the characteristics of Bitcoin and Ethereum.
It introduces an Account Abstraction Layer to integrate Bitcoin's Unspent Transaction Output model with the Ethereum Virtual Machine for smart contracts to operate~\cite{QTUMEXP}.
Besides, Qtum adopts Proof-of-Stake (PoS) consensus mechanism instead of Bitcoin's Proof-of-Work (PoW) to simplify the mining process since PoW is resource-intensive, i.e., it wastes enormous amounts of electricity on mining coins~\cite{QTUMPOS}.

\textbf{Ravencoin} is unique in terms of that it was designed for users to tokenize assets on-chain and transfer ownership via blockchain transactions~\cite{RVNPaper}.
Such assets can be physical or digital, including gold, in-game items, copyrights, etc~\cite{RVNVSBTC}.

\subsection{Ethereum and its Forked Projects}
\label{subsec:ethereum_forks}

\textbf{Ethereum}~\cite{ETHYellowPaper2022} is the first blockchain system with the capability of constructing Turing-complete \textit{smart contracts}, which contain a set of pre-defined rules and regulations for self-execution.
Ether (ETH) is the native cryptocurrency for maintaining the operations on Ethereum, which is the second largest cryptocurrency with a market capitalization of around 230 billion USD as of June 2022.
As an open-sourced project, Ethereum also nourished many blockchain projects.
Specifically, we analyzed all the projects listed on Blockscan~\cite{BSCAN} and selected five of the most popular projects that directly fork or partially reuse the code of \eth.
\mytab~\ref{tab:repo_info_eth} presents the basic information of these forked projects as of 6 June 2022.

\textbf{Binance} is the largest cryptocurrency exchange in the world.
As of 27 July 2022, its 24-hour trading volume reaches 11.7 billion USD~\cite{BNBEXC}.
Originally, Binance developed Binance Chain to provide a marketplace for trading cryptocurrency in a decentralized manner, with BNB being the native token.
However, as Binance Chain is not EVM-compatible, users cannot develop decentralized applications (DApps) using smart contracts~\cite{BCVSBSC}.
Binance initiated Binance Smart Chain (BSC) with EVM compatibility to solve this problem.
On February 15, 2022, Binance Chain and Binance Smart Chain united into BNB Chain~\cite{BNBCHAIN}.
Currently, BNB Chain holds around 3.4 million transactions daily, with 2.0 million active wallets~\cite{BNBSCAN}.

\textbf{Avalanche} aims to solve \eth's issues regarding transaction fee, scalability, and programmability, by leveraging a multi-chain approach~\cite{AVAXVSETH}.
Specifically, Avalanche combines three separate blockchain networks, i.e., X-Chain: for issuing digital assets, C-Chain: for converting \eth's DApps to Avalanche, and P-Chain: for validating the states of subnets.
\textbf{Celo} is also EVM-compatible.
Notably, it provides a client designed for mobile phone users.
Moreover, while the transaction fee is paid with the native asset (ETH) on \eth, Celo allows users to pay transaction fees with the native asset (CELO) and stable coins (cUSD and cEUR)~\cite{CELOVSETH}.

\textbf{Polygon} and \textbf{Optimism} are \eth's layer-2 networks, which also target on \eth's scalability and transaction fee issues.
Layer-2 solutions refer to infrastructures or simple protocols built on top of the \eth main chain~\cite{POLYVSOP}, i.e., layer-1.
Typically, they handle off-chain transactions and send only compact data to layer-1.
Polygon is technically a sidechain of \eth, as it uses its own consensus algorithms and runs in parallel with the main chain.
However, different from sidechains, Optimism uses Optimistic Rollups~\cite{OPROLL} to interact with the main chain and use smart contracts that reside within \eth~\cite{LAYER2}.

\begin{figure*}[t!]
	\begin{adjustbox}{center}
		\includegraphics[width=1.0\linewidth]{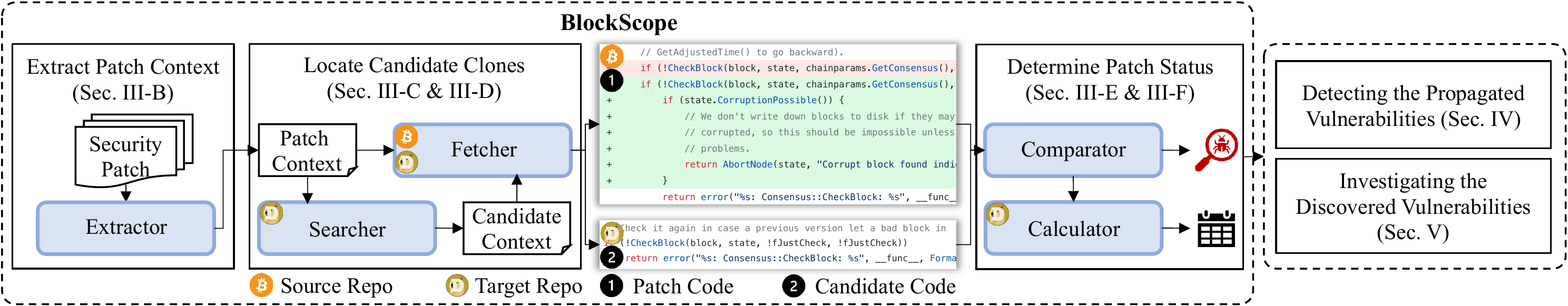}
	\end{adjustbox}
    \caption{The overall workflow of \name and our study.}
	\vspace{-1ex}
	\label{fig:workflow}
\end{figure*}

\subsection{Definition of Code Clone Types}
\label{subsec:clone_types}

Due to the nature of open-source projects, it is common for projects to reuse parts of code from others.
However, vulnerabilities are always reintroduced due to the casual code reuses, namely code clones.
While code clone detections are widely studied among the famous open-source projects, e.g., Linux Kernel, detections for cloned vulnerabilities in the forked blockchain projects are much less explored.
In this study, it is essential to analyze the cloned code among the forked blockchain projects. 
Therefore, we adopt the type definitions of code clones from \cite{CLONEDEF} as follows:

\begin{compactitem}
\item Type-1 clones refer to two identical code fragments with variations in whitespaces, layouts, and comments.
\item Type-2 clones include Type-1 clones and extend the variations to identifiers, literals, and types, e.g., variable renaming.
\item Type-3 clones further extend these variations to syntactically similar code with inserted, deleted, or updated statements.
\item Type-4 clones refer to semantically equivalent code fragments but syntactically different, which is out of the scope of this paper.
\end{compactitem}

In this paper, we focus on the detection of Type-1, Type-2, and Type-3 code clones.
Detecting Type-4 code clones requires code semantic learning or understanding, which is out of the scope of typical clone detection tools including \name.



%% file: resources/tab/tab_repo_info.tex
\begin{table}[t!]
	\centering
	\vspace{-2ex}
	\caption{The basic information of Bitcoin, Ethereum, and their popular forked projects.}
	\label{tab:repo_info}
	\vspace{-3ex}
	\begin{subtable}{1.0\linewidth}
		\centering
		\caption{Bitcoin and its forked projects (as of 7 September 2021).}
		\label{tab:repo_info_bit}
		\begin{adjustbox}{center} 
		\scalebox{0.97}{
			\begin{tabular}{|c|c|c|c|l|c|}
				\hline
				\textbf{\#} & \textbf{Name} & \textbf{Code} & \textbf{Market Cap} & \textbf{Repository} & \textbf{Star} \\ \hline
				1   & Bitcoin      & BTC  & \$749.70B & bitcoin/bitcoin           & 60.3K \\
				6   & Dogecoin     & DOGE & \$42.55B  & dogecoin/dogecoin         & 13.6K \\
				11  & Bitcoin Cash & BCH  & \$12.02B  & Bitcoin-ABC/bitcoin-abc   & 1.1K  \\
				12  & Litecoin     & LTC  & \$11.88B  & litecoin-project/litecoin & 4K    \\
				33  & Bitcoin SV   & BSV  & \$3.24B   & bitcoin-sv/bitcoin-sv     & 520   \\
				55  & Dash         & DASH & \$1.79B   & dashpay/dash              & 1.4K  \\
				59  & Zcash        & ZEC  & \$1.64B   & zcash/zcash               & 4.5K  \\
				75  & Bitcoin Gold & BTG  & \$1.04B   & BTCGPU/BTCGPU             & 611   \\
				79  & Horizen      & ZEN  & \$935.27M & HorizenOfficial/zen       & 202   \\
				80  & Qtum         & QTUM & \$923.88M & qtumproject/qtum          & 1.1K  \\
				83  & DigiByte     & DGB  & \$868.91M & digibyte/digibyte         & 361   \\
				100 & Ravencoin    & RVN  & \$693.34M & RavenProject/Ravencoin    & 932   \\ \hline
			\end{tabular}%
		}
		\end{adjustbox}
	\end{subtable}
	\vfill
	\begin{subtable}{1.0\linewidth}
		\centering
		\caption{Ethereum and its forked projects (as of 6 June 2022).}
		\label{tab:repo_info_eth}
		\begin{adjustbox}{center} 
		\scalebox{0.95}{
			\begin{tabular}{|c|c|c|c|l|c|}
				\hline
				\textbf{\#} & \textbf{Name} & \textbf{Code} & \textbf{Market Cap} & \textbf{Repository} & \textbf{Star} \\ \hline
				2   & Ethereum      & ETH  & \$229.87B & ethereum/go-ethereum           & 37.7K \\
				5   & Binance   & BNB & \$50.69B  & bnb-chain/bsc         & 1.6K \\
				14  & Avalanche & AVAX  & \$7.65B  & ava-labs/subnet-evm   & 1.6K  \\
				17  & Polygon     & MATIC  & \$5.15B  & maticnetwork/bor & 400    \\
				78  & Celo         & CELO & \$604.02M   & celo-org/celo-blockchain              & 382  \\
				199 & Optimism   & OP  & \$263.36M   & ethereum-optimism/optimism     & 1.2K   \\ \hline
			\end{tabular}%
		}
		\end{adjustbox}
	\end{subtable}
\end{table}

%% file: method.tex

\section{\name}
\label{sec:method}

\input{overview}
\subsection{Extracting Patch Contexts from the Source Repositories}
\label{subsec:manual_extract}




Given a security patch from the source project or code repository (e.g., Bitcoin/\eth), \name first extracts its code context.
In this paper, we provide an \texttt{Extractor} component to \textit{automatically} extract the contexts of patch code and use its output for system evaluation.
In reality, \name also supports the manually crafted code contexts from security experts for better accuracy.
To distinguish the context of patch code from that of target code, we call the former ``\textit{patch context}'' and the latter ``\textit{candidate context}'', as shown in \myfig~\ref{fig:workflow}.


\begin{figure*}[t!]
	\begin{adjustbox}{center}
		\includegraphics[width=1.0\linewidth]{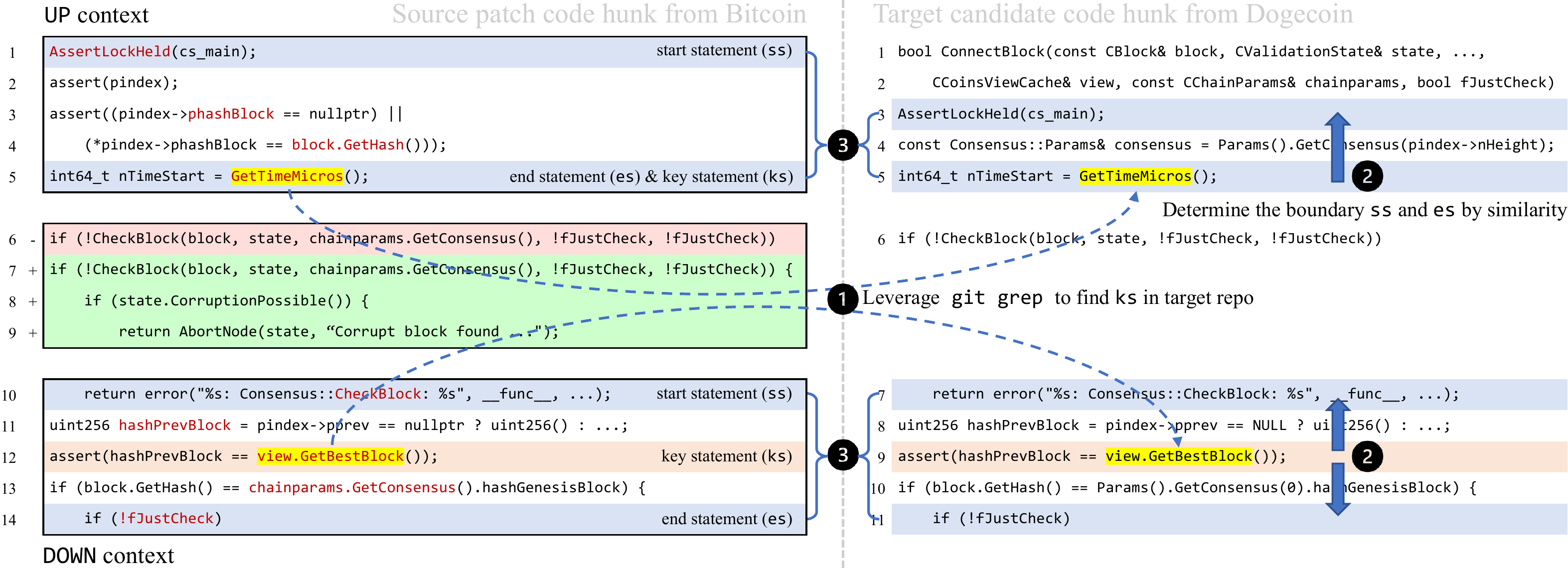}
	\end{adjustbox}
    \caption{Illustrating \name's context-based search process for finding candidate contexts in a target repository.} 
	\vspace{-1ex}
	\label{fig:searcher}
\end{figure*}

Unlike ReDeBug that directly takes the entire part of the nearby code lines (after normalization and tokenization) as context, \texttt{Extractor} recognizes important variable and function names as the \textit{context keywords} and uses these keywords to search for candidate contexts in the target repositories (as in \mysec\ref{subsec:search_code}).
As a result, we do not require each extracted keyword to be precise because \textit{as long as one of the context keywords can find the correct candidate context}, context similarity measurement (in \mysec\ref{subsec:search_code}) will automatically exclude the search results of other incorrect context keywords.

We use the left patch code of \myfig~\ref{fig:searcher} to illustrate the process of extracting context keywords. 
After normalizing and tokenizing each patch code line, \texttt{Extractor} uses the following heuristics to automatically recognize at most one context keyword per code line.
Specifically, we consider the tokens with both lower and upper case letters (including some special characters like ``.'') and select the longest one as the most important variable or function name of one code line.
In this way, \name automatically selects nine context keywords, as highlighted in red color, from the patch code context in \myfig~\ref{fig:searcher}.
As mentioned above, we do not require each extracted keyword to be precise, and according to our evaluation in \mysec\ref{sec:detect}, this simple strategy of automatically extracting context keywords works well for our problem. 

\subsection{Searching for Candidate Contexts in the Target Repositories}
\label{subsec:search_code}

The \texttt{Searcher} component of \name then uses the extracted context keywords to search for candidate contexts in the target repositories.
The basic idea is to first search for the key statements in target code (via patch context keywords), then recover the corresponding boundary of each potential code context, and finally determine the candidate contexts via the similarity measurement with the original patch context.
To illustrate this context-based search process, we use Bitcoin's patch of checking corrupted blocks and its vulnerable clone in Dogecoin as a running example.
As shown in \myfig~\ref{fig:searcher}, the left-hand side is the patch code hunk (commit \texttt{0e7c52dc}) from Bitcoin, while the right-hand side shows the cloned version in Dogecoin \footnote{Note that we only keep the ``meaningful'' code statements, i.e., empty lines, comments, and single brackets are removed.}.
It also illustrates the following three steps.

\textbf{1) Searching for the key statements.} 
The first step is to find the key statements (\texttt{ks}) that are the code statements in the target code with the searched context keywords.
Specifically, \texttt{Searcher} first leverages \texttt{git grep} to search for all the code statements that contain the patch context keyword(s) in the target repositories, and then finalize the search result by measuring the similarity between the searched \texttt{ks} with the original \texttt{ks}.
If the measured similarity is higher than the threshold configured in \name, we consider it one potential candidate \texttt{ks}.
To minimize the misses and avoid causing false negatives to the subsequent steps, this step uses a relatively low threshold (0.25) based on the Normalized Levenshtein~\cite{TPAMI07Levenshtein} metric, i.e., $\textup{strsim}()$ used in equation~\eqref{eq:ksim}. 
This is acceptable because 
among all the searched candidate \texttt{ks}s, we select the one with the highest similarity as the final candidate \texttt{ks}.
Specifically, given a patch context $pc = \{(k_{1}, s_{1}), (k_{2}, s_{2}), ..., (k_{m}, s_{m})\}$, where $(k_{i}, s_{i})$ represents the extracted keyword $k_{i}$ of the code statement $s_{i}$, the search result $sr_{i}$ for $k_{i}$ is represented as $sr_{i} = (k_{i}, [s'_{i1}, s'_{i2}, ..., s'_{in}])$, where $s'_{ij}$ is the code statement that contains $k_{i}$ in the target repository.
We determine $s'_{pq}$ as the final candidate \texttt{ks} according to the equation~\eqref{eq:ksim}.
In the case of \myfig~\ref{fig:searcher}, \texttt{Searcher} selects line 5 and 9 (both with the highest similarity) of Dogecoin as the final candidate \texttt{ks}s of the \texttt{UP} and \texttt{DOWN} contexts, respectively.

\begin{equation} \label{eq:ksim}
p, q = \mathop{\arg\max}_{1\leq i \leq m, 1\leq j \leq n} \ \textup{strsim}(s_{i},s'_{ij})
\end{equation}

Moreover, in the course of implementing the candidate context search, we adopt three \textit{automatic} optimizations to further improve \name's context search precision and avoid unnecessary analysis in the subsequent steps.
First, it excludes the search result with comments and test code.
Second, it excludes the search result with the file type different from the patch's file type, e.g., the patch in \myfig~\ref{fig:searcher} is a C/C++ source code file, based on which \name excludes C/C++ header files and non-C/C++ source code files in the search result.
Third, \name excludes the search result with different statement types.
For example, since line 5 in \myfig~\ref{fig:searcher} is an assignment statement, any search result does not match the same statement type will be automatically discarded.



\textbf{2) Determining the boundary of candidate contexts.}
Once identified the candidate \texttt{ks}, the next step of \texttt{Searcher} is to retrieve the code statements surrounding it and determine their boundary.
Specifically, we need to expand the one-line candidate \texttt{ks} into the multi-line candidate context that has the corresponding boundary as the original patch context.
To do so, we first fetch the same number of nearby code statements from target code as that, represented as \texttt{C\_LINES}, in the patch context.
For example, in \myfig~\ref{fig:searcher}, if we set \texttt{C\_LINES=5}, 
\texttt{Searcher} fetches line 1 to 5 and line 7 to 11 for the candidate \texttt{UP} and \texttt{DOWN} contexts in Dogecoin, respectively.
Then starting from the \texttt{ks} (i.e., line 5 and 9 of Dogecoin), \texttt{Searcher} compares each code statement upwards and downwards with the start statement (\texttt{ss}) and end statement (\texttt{es}) in the patch context, respectively.
It then selects the ones with the highest similarity and also exceeding the aforementioned threshold (0.25) as the boundary \texttt{ss} and \texttt{es} in the candidate context, e.g., line 3 and line 5 for Dogecoin's \texttt{UP} context.

\textbf{3) Finalizing the candidate contexts via similarity measurement.}
It is worth noting that \texttt{ss} and \texttt{es} only define the boundary of the candidate context, while the code statements in between remain unchecked.
As illustrated in the step 3 of \myfig\ref{fig:searcher}, we thus further check whether the entire candidate context is indeed similar to the patch context via the same multi-line code similarity measurement that will be introduced in \mysec\ref{subsec:code_sim}.
If the measured similarity between the candidate context \textit{C} and the patch context \textit{P} exceeds a threshold, we consider \textit{C} as the context of a candidate clone for further processing; otherwise, we discard this candidate context.
Note that since multiple candidate contexts' similarity could exceed the threshold, all of these candidate contexts will be further processed.


\subsection{Fetching Patch and Candidate Code Hunks from the Source and Target Repositories}
\label{subsec:fetch_code}

With the determined candidate context(s), we leverage \texttt{Fetcher} to retrieve the patch code from the source repository and the candidate code from the target repository, respectively.
Note that \texttt{Fetcher} is also used by the earlier \texttt{Searcher} component to retrieve the context of a patch/candidate code hunk. 
Specifically, a typical code hunk consists of three code fragments, the \texttt{UP} context, the \texttt{DOWN} context, and the middle patch/candidate code, as previously shown in \myfig~\ref{fig:searcher}.

For the patch code hunk, \texttt{Fetcher} directly fetches its patch code from the commit history and selects the nearby code statements upwards and downwards (with the line number specified by \texttt{C\_LINES}) as the \texttt{UP} and \texttt{DOWN} contexts, respectively.
For the candidate code hunk, we fetch its code statements according to the candidate context determined in \mysec\ref{subsec:search_code} and also the original patch context.
Specifically, if the original patch contains both \texttt{UP} and \texttt{DOWN} contexts, we regard the code statements between the corresponding candidate contexts as the candidate code.
As a result, line 6 of Dogecoin is fetched as the candidate code in \myfig~\ref{fig:searcher}.
If the patch context contains only the \texttt{UP} context, we regard the code statements below it as the candidate code.
Similarly, if the patch context contains only the \texttt{DOWN} context, we regard the code statements above it as the candidate code.
Note that for the last two situations, the candidate code is fetched with the same number of code statements as the patch code.

\subsection{Measuring the Similarity between Patch and Candidate Code}
\label{subsec:code_sim}

With the fetched patch and candidate code, \texttt{Comparator} measures the similarity between their two code fragments and also determine whether the target repository has fixed the vulnerability, if the candidate code is not vulnerable.
As mentioned in \mysec\ref{subsec:overview}, we need a new way of calculating the code similarity that is immune to Type-1/2/3 clones.

We first abstract the code similarity problem in this form: given a source code fragment \textit{S} with \textit{p} code statements and a target code fragment \textit{T} with \textit{q} code statements, respectively, we need to design an appropriate measure to determine their similarity.
Intuitively, we can compute the similarity between \textit{S} and \textit{T} by first adding up the similarity of each pair of code statements at the same position in \textit{S} and \textit{T} and then normalizing it into $[0,1]$, i.e., $\frac{1}{p}\sum_{i=1}^{p}\textup{strsim}(S_{i},T_{i})$.
While this can handle Type-2 clones because of not using the hash-based exact match per code line, it is still not applicable to measuring Type-3 clones for two reasons.
First, as Type-3 clones involve inserted/deleted statements, i.e., $p\neq q$, the extra code statements will not be measured in this way.
Second, because of the inserted/deleted statements, the ordering of the same code statement in \textit{S} and \textit{T} might be also different.

To solve the problems above, we determine two principles: (i) all the code statements in \textit{S} and \textit{T} should be considered; and (ii) the influence of the ordering issue should be adjustable.
For the first principle, we identify the most similar code statement in \textit{T} for every code statement in \textit{S}, i.e., for each code statement $S_{i}\in S$, we find $T_{j}\in T$, s.t., $j=\mathop{\arg\max}_{k} \ \textup{strsim}(S_{i},T_{k})$. 
For the second principle, we first define the index \textit{i} and \textit{j} as the relative positions of the code statements in \textit{S} and \textit{T} if $S_{i}$'s most similar statement is $T_{j}$.
The basic idea is that the greater the difference between \textit{i} and \textit{j} is, the less similarity between $S_{i}$ and $T_{j}$ should be.
Therefore, we introduce a parameter $r\in [0,1]$, and $r^{|i-j|}$ to indicate the reward of the similarity between $S_{i}$ and $T_{j}$.
By multiplying this reward by the original similarity, we can adjust the ordering issue's influence on code similarity.
To illustrate the impact of $r$ on the similarity measurement, we calculate the similarities of all the patch and candidate code pairs under different $r$. We present the result in Appendix \ref{sec:appendix1}.
In this paper, we set 0.95 as the default value of $r$.
Once finishing the calculation of such similarity for every code statement in \textit{S}, we sum them up and normalize the result into $[0,1]$, as shown in the following equation~\eqref{eq:sim}.

\vspace{-3ex}
\begin{equation} \label{eq:sim}
\begin{aligned}
\textup{SIMILARITY}(S,T) &= \frac{1}{p}\sum_{i=1}^{p} \textup{strsim}(S_{i},T_{j})r^{|i-j|} \\
\textup{s.t.,} \ j &= \mathop{\arg\max}_{1\leq k \leq q} \ \textup{strsim}(S_{i},T_{k})
\end{aligned}
\end{equation}
\vspace{-1ex}

While the method above provides a new way of measuring the similarity between two code fragments, we still need to determine whether the target repository has applied a patch or not.
Specifically, given the candidate code \textit{C} of the target repository, we compare it with the patch code \textit{P}.
Note that there are three types of \textit{P}:
(i) \texttt{DEL}-type: contains only the deleted lines, i.e., $P = [dp]$;
(ii) \texttt{ADD}-type: contains only the added lines, i.e., $P = [ap]$;
and (iii) \texttt{CHA}-type: contains both deleted and added lines, i.e., $P = [dp, ap]$.
We thus determine the comparison logic as follows (where $t$ is the threshold):

\begin{itemize}
  \item For type (i), if $\textup{SIMILARITY}(C, dp) \geq t$, we determine that \textit{C} did \textit{not} apply \textit{P}; otherwise, we determine that \textit{C} has applied \textit{P}.

  \item For type (ii), if $\textup{SIMILARITY}(C, ap) \geq t$, we determine that \textit{C} has applied \textit{P}; otherwise, we determine that \textit{C} did \textit{not} apply \textit{P}.

  \item For type (iii), if $\textup{SIMILARITY}(C, dp) \geq t$ and $\textup{SIMILARITY}(C, ap) \geq t$ and $\textup{SIMILARITY}(C, dp) \geq \textup{SIMILARITY}(C, ap)$, we determine that \textit{C} did \textit{not} apply \textit{P}; otherwise, if $\textup{SIMILARITY}(C, dp) \geq t$ and $\textup{SIMILARITY}(C, ap) \geq t$ and $\textup{SIMILARITY}(C, dp) < \textup{SIMILARITY}(C, ap)$, we determine that \textit{C} has applied \textit{P}.
\end{itemize}

Moreover, as \texttt{Searcher} may return multiple candidate contexts in the target repository, leading to multiple candidate code, i.e., $C_{i} \in [C_{1}, C_{2}, ..., C_{n}]$.
For each $C_{i}$, we calculate $s_{i} = \textup{SIMILARITY}(C_{i}, P)$, and determine its patch applying status $fv_{i} \in \{0, 1\}$, where $fv_{i} = 1$ $(= 0)$ indicates $C_{i}$ has (not) applied \textit{P}.
Here we introduce a factor $conf_{i}$ to measure the confidence of $fv_{i}$ on $C_{i}$ by $conf_{i} = s_{i} - t$, i.e., the greater $s_{i}$ exceeds $t$ the more confident $fv_{i}$ is on $C_{i}$.
Finally, we can determine the status of \textit{P} in the target repository by the most confident $fv_{i}$, i.e., $i = \mathop{\arg\max}_{j} conf_{j}$.
If the target repository did not apply \textit{P}, we consider it a vulnerability; otherwise, we consider the vulnerability fixed.


\input{calculator}

%% file: overview.tex

\subsection{Design Choices and System Overview}
\label{subsec:overview}

To detect the propagated vulnerabilities from the existing security patches of Bitcoin/\eth, we design \name as a patch-based code clone detection tool.
This makes \name, by nature, more similar to security-oriented clone detection tools (e.g., ReDeBug~\cite{Redebug2012}, VUDDY~\cite{VUDDY2017}, MVP~\cite{MVP2020}, and VGraph~\cite{bowmanvgraph}) rather than the traditional clone detection tools (e.g., CCFinder~\cite{CCFinder02}, CPMiner~\cite{CPMiner04}, DECKARD~\cite{deckard_2007}, and SourcererCC~\cite{Sourcercc2016}) that do not differentiate vulnerable and patched code inputs.
Moreover, since we aim to test all different blockchain projects, we design \name to be language-agnostic as similar to ReDeBug.
As a result, we do not perform ``program analysis-alike'' preprocessing, such as variable/type/function abstraction in VUDDY, program slicing in MVP, and code property graph~\cite{CodePropertyGraph14} in VGraph,  before the similarity measurement between source and target code.

Besides the choices above, \name offers two unique designs that are also the major novelty of our methodology:
\begin{compactitem}

	\item \textit{Leveraging patch code contexts to search and locate only potentially relevant code.}
    Since our detection targets are the propagated vulnerabilities in the forked projects, it is reasonable to assume that they have similar contexts as the original patch code in the source repositories.
    \name thus leverages the extracted patch code contexts to search for potentially relevant code in the target repositories and employs code similarity to finalize the contexts of candidate code clones.
    This not only helps \name avoid the whole-repository analysis as in typical code clone detection tools but also improves the precision because the context similarity is also being considered. 

	\item \textit{Adopting similarity-based code match for being more tolerant to variant code clones.}
    To cover all the syntax-wide Type-1, Type-2, and Type-3 clones, we adopt similarity-based code match, instead of the hash-based exact code match in ReDeBug~\cite{Redebug2012}, VUDDY~\cite{VUDDY2017}, and MVP~\cite{MVP2020}.
    This allows \name to be more tolerant to the code lines with no exact ``abstracted'' hashes (i.e., Type-2 clones).
    Moreover, we design a new way of calculating code similarity to better handle the code fragments with inserted/deleted/reordered code lines (i.e., Type-3 clones). 

\end{compactitem}

\myfig~\ref{fig:workflow} presents the overall workflow of \name in five major steps.
Firstly, \mysec\ref{subsec:manual_extract} describes how the \texttt{Extractor} component or \texttt{Extractor}\footnote{We describe different \name components using their names, e.g., \texttt{Extractor}, hereafter.} extracts the code contexts from patches in the source repositories.
Secondly, in \mysec\ref{subsec:search_code}, \texttt{Searcher} leverages the extracted patch contexts to search for candidate contexts in the target repositories.
Thirdly, \texttt{Fetcher} in \mysec\ref{subsec:fetch_code} retrieves the patch and candidate code hunks in the source and target repositories, respectively.
Fourthly, \texttt{Comparator} in \mysec\ref{subsec:code_sim} employs a new similarity-based code matching technique to determine the propagated vulnerabilities from \texttt{Fetcher}'s outputs.
Lastly, for the vulnerabilities already patched, \texttt{Calculator} in \mysec\ref{subsec:release_delay} measures their patch delays in the target repositories.

%% file: calculator.tex
\subsection{Determining Patch Delays for the Vulnerabilities Already Patched in the Target Repositories}
\label{subsec:release_delay}

For the vulnerabilities already patched in the target repositories, we further leverage \texttt{Calculator} to automatically measure their patch delays.
We define the \textit{patch delay} as the interval between the patch's commit date in the source project and the patch's release date in the target project because eventually, the release date is the actual time when a patch is available to the blockchain node operators and end users.

\input{resources/tab/tab_blame_example.tex}

Upon receiving a candidate code that is determined as fixed, \texttt{Calculator} leverages \texttt{git blame} to retrieve the commit that patched the code.
\mytab~\ref{tab:blame_exp} illustrate an example output of \texttt{git blame}, where the left column shows the commit hash (SHA), the column in the middle shows the line number for the code statements on the right in Qtum's \texttt{src/qt/bitcoin.cpp} file.
The code from line 204 to line 208 is actually Qtum's patch for fixing the cloned CVE-2021-3401~\cite{CVE-2021-3401-Patch-Code} in its project.
It was added by two commits, \texttt{a2714a5c69} and \texttt{797fef7bee}, where \texttt{797fef7bee} only modified line 205.
Hence, we still need to determine which commit is the \textit{true} fix. 
In the Qtum example, after checking both commits, we identify that line 205 in \mytab~\ref{tab:blame_exp} was originally added by \texttt{a2714a5c69} on 10 August 2019 as \texttt{static const char* qt\_argv = "bitcoin-qt";}, where \texttt{"bitcoin-qt"} is later replaced by \texttt{"qtum-qt"} in \texttt{797fef7bee} on 26 June 2020.
As a result, if multiple commits modify the candidate code, we consider the earliest one is the \textit{true} fix commit.


Moreover, we need to scrape the release information from GitHub because the local git repository does not contain such information.
By analyzing a commit's GitHub webpage, \texttt{Calculator} can retrieve all of its release versions and determine the earliest date when the commit was first released.
In the Qtum example, the patch commit \texttt{a2714a5c69} was first released in the version \texttt{mainnet-ignition-v0.19.0} on 22 February 2020, which was delayed from the original Bitcoin commit by 197 days.

%% file: resources/tab/tab_blame_example.tex
\begin{table}[t!]
	\centering
	\vspace{-1ex}
	\caption{An example of the output of \texttt{git blame}.}
	\label{tab:blame_exp}
	\vspace{-1ex}
	\begin{adjustbox}{center} 
	\scalebox{0.85}{
		\begin{tabular}{|crl|}
			\multicolumn{3}{l}{src/qt/bitcoin.cpp} \\
			\hline
			202d853b & 201 & \texttt{\hspace{5mm}\}}                       \\
			202d853b & 202 & \texttt{\}}                       \\
			202d853b & 203 &                                   \\
			a2714a5c & 204 & \texttt{static int qt\_argc = 1;} \\
			797fef7b & 205 & \texttt{static const char* qt\_argv = "qtum-qt";}                          \\
			a2714a5c & 206 &                                   \\
			a2714a5c & 207 & \texttt{BitcoinApplication::BitcoinApplication(...):} \\
			a2714a5c & 208 & \texttt{\hspace{5mm}QApplication(qt\_argc, const\_cast<char **>(...)),}        \\
			9096276e & 209 & \texttt{\hspace{5mm}coreThread(nullptr),}     \\
			71e0d908 & 210 & \texttt{\hspace{5mm}m\_node(node),}            \\
			9096276e & 211 & \texttt{\hspace{5mm}optionsModel(nullptr),}   \\ \hline
		\end{tabular}%
	}
	\end{adjustbox}
\end{table}

%% file: detect.tex
\section{Detecting the Vulnerabilities Propagated to Forked Projects} 
\label{sec:detect}

In this section, we aim to detect the vulnerabilities that are propagated from Bitcoin and Ethereum to their forked blockchain projects using \name.
To this end, we first benchmark the accuracy and performance of \name (\mysec\ref{sec:accuracy}) using an experimental setup introduced in \mysec\ref{sec:setup}.
We then present the detected vulnerabilities in \mysec\ref{sec:result}.
Finally, we conduct ethical vulnerability reporting and summarize vendors' response/actions in \mysec\ref{sec:report}.

\input{resources/tab/tab_exp_result}

\input{setup}

\input{accuracy}
\input{result}
\input{report}

%% file: resources/tab/tab_exp_result.tex
\begin{table*}[t!]
	\centering
    \vspace{-1ex}
	\caption{The experimental result of \name.}
    \vspace{-2ex}
	\label{tab:exp_result}
	\begin{subtable}[t!]{0.67\linewidth}
		\centering
		\caption{The accuracy and performance comparison between \name and ReDeBug.}
		\label{tab:exp_compare}
		\begin{adjustbox}{center} 
		\scalebox{1}{
		\begin{threeparttable}
			\begin{tabular}{|c|cccccc|ccccc|}
				\hline
				\multirow{2}{*}{\textbf{Forked Project}} &
				  \multicolumn{1}{c|}{\multirow{2}{*}{\textbf{LOC}}} &
				  \multicolumn{5}{c|}{\textbf{\name}} &
				  \multicolumn{5}{c|}{\textbf{ReDeBug}} \\ \cline{3-12} 
				 &
				  \multicolumn{1}{c|}{} &
				  \textbf{TP} &
				  \textbf{FN} &
				  \textbf{TN} &
				  \multicolumn{1}{c|}{\textbf{FP}} &
				  \textbf{Time} &
				  \textbf{TP} &
				  \textbf{FN} &
				  \textbf{TN} &
				  \multicolumn{1}{c|}{\textbf{FP}} &
				  \textbf{Time} \\ \hline
				Dogecoin     & \multicolumn{1}{c|}{326.9K}            & 16 & - & 15 & \multicolumn{1}{c|}{1} & 7.6s            & 7  & 9  & 15 & \multicolumn{1}{c|}{1} & 12.5s            \\
				Bitcoin Cash & \multicolumn{1}{c|}{607.1K}            & 1  & - & 30 & \multicolumn{1}{c|}{1} & 10.5s           & -  & 1  & 31 & \multicolumn{1}{c|}{-} & 22.2s            \\
				Litecoin     & \multicolumn{1}{c|}{423.3K}            & 6  & - & 26 & \multicolumn{1}{c|}{-} & 8.3s            & 5  & 1  & 26 & \multicolumn{1}{c|}{-} & 16.4s            \\
				Bitcoin SV   & \multicolumn{1}{c|}{221.1K}            & 11 & 1 & 18 & \multicolumn{1}{c|}{2} & 10.6s           & 2  & 10 & 19 & \multicolumn{1}{c|}{1} & 9.9s             \\
				Dash         & \multicolumn{1}{c|}{380.3K}            & 9  & 1 & 22 & \multicolumn{1}{c|}{-} & 13.9s           & 7  & 3  & 21 & \multicolumn{1}{c|}{1} & 17.7s            \\
				Zcash        & \multicolumn{1}{c|}{199.4K}            & 9  & 2 & 19 & \multicolumn{1}{c|}{2} & 8.4s            & 1  & 10 & 21 & \multicolumn{1}{c|}{-} & 10.7s            \\
				Bitcoin Gold & \multicolumn{1}{c|}{381.7K}            & 10 & 1 & 21 & \multicolumn{1}{c|}{-} & 8.8s            & 10 & 1  & 21 & \multicolumn{1}{c|}{-} & 17.4s            \\
				Horizen      & \multicolumn{1}{c|}{178.9K}            & 9  & 2 & 20 & \multicolumn{1}{c|}{1} & 7.7s            & 1  & 10 & 21 & \multicolumn{1}{c|}{-} & 12.6s            \\
				Qtum         & \multicolumn{1}{c|}{569.0K}            & -  & - & 31 & \multicolumn{1}{c|}{1} & 12.0s           & -  & -  & 32 & \multicolumn{1}{c|}{-} & 33.5s            \\
				DigiByte     & \multicolumn{1}{c|}{416.3K}            & 10 & 1 & 21 & \multicolumn{1}{c|}{-} & 10.7s           & 10 & 1  & 21 & \multicolumn{1}{c|}{-} & 15.8s            \\
				Ravencoin    & \multicolumn{1}{c|}{504.2K}            & 14 & 1 & 16 & \multicolumn{1}{c|}{1} & 11.4s           & 10 & 5  & 17 & \multicolumn{1}{c|}{-} & 20.9s            \\ \hline
				\multirow{2}{*}{\textbf{Sum}} &
				  \multicolumn{1}{c|}{\textbf{4.2M}} &
				  \multirow{2}{*}{\textbf{95}} &
				  \multirow{2}{*}{\textbf{9}} &
				  \multirow{2}{*}{\textbf{239}} &
				  \multicolumn{1}{c|}{\multirow{2}{*}{\textbf{9}}} &
				  \textbf{109.9s} &
				  \multirow{2}{*}{\textbf{53}} &
				  \multirow{2}{*}{\textbf{51}} &
				  \multirow{2}{*}{\textbf{245}} &
				  \multicolumn{1}{c|}{\multirow{2}{*}{\textbf{3}}} &
				  \textbf{189.6s} \\
				             & \multicolumn{1}{c|}{\textbf{(382.6K)}*} &    &   &    & \multicolumn{1}{c|}{}  & \textbf{(3.4s)}$^{\diamond}$ &    &    &    & \multicolumn{1}{c|}{}  & \textbf{(5.9s)}$^{\diamond}$  \\ \hline \hline
				Binance      & \multicolumn{1}{c|}{565.3K}            & 1  & - & 5  & \multicolumn{1}{c|}{-} & 2.2s            & -  & 1  & 5  & \multicolumn{1}{c|}{-} & 30.2s            \\
				Avalanche    & \multicolumn{1}{c|}{1070.1K}           & -  & - & 6  & \multicolumn{1}{c|}{-} & 2.5s            & -  & -  & 6  & \multicolumn{1}{c|}{-} & 55.2s            \\
				Polygon      & \multicolumn{1}{c|}{592.0K}            & -  & - & 6  & \multicolumn{1}{c|}{-} & 2.3s            & -  & -  & 6  & \multicolumn{1}{c|}{-} & 31.3s            \\
				Celo         & \multicolumn{1}{c|}{631.0K}            & 1  & - & 5  & \multicolumn{1}{c|}{-} & 2.7s            & 1  & -  & 5  & \multicolumn{1}{c|}{-} & 44.5s            \\
				Optimism     & \multicolumn{1}{c|}{630.6K}            & 4  & - & 2  & \multicolumn{1}{c|}{-} & 3.6s            & 3  & 1  & 2  & \multicolumn{1}{c|}{-} & 43.3s            \\ \hline
				\multirow{2}{*}{\textbf{Sum}} &
				  \multicolumn{1}{c|}{\textbf{3.5M}} &
				  \multirow{2}{*}{\textbf{6}} &
				  \multirow{2}{*}{\textbf{-}} &
				  \multirow{2}{*}{\textbf{24}} &
				  \multicolumn{1}{c|}{\multirow{2}{*}{\textbf{-}}} &
				  \textbf{13.3s} &
				  \multirow{2}{*}{\textbf{4}} &
				  \multirow{2}{*}{\textbf{2}} &
				  \multirow{2}{*}{\textbf{24}} &
				  \multicolumn{1}{c|}{\multirow{2}{*}{\textbf{-}}} &
				  \textbf{204.5s} \\
				             & \multicolumn{1}{c|}{\textbf{(697.8K)}*}  &    &   &    & \multicolumn{1}{c|}{}  & \textbf{(2.2s)}$^{\diamond}$ &    &    &    & \multicolumn{1}{c|}{}  & \textbf{(34.1s)}$^{\diamond}$ \\ \hline
			\end{tabular}
			\begin{tablenotes}
				\item \small *: the numbers in (.) of these cells represent the average LOC per \textit{project}.
				\item \small $^{\diamond}$: the numbers in (.) of these cells represent the average processing time per \textit{patch}.
			\end{tablenotes}
		\end{threeparttable}
		}
		\end{adjustbox}
	\end{subtable}
	\hfill
	\begin{subtable}[t!]{0.32\linewidth}
		\centering
		\caption{The fixed cases detected by \name.}
		\label{tab:fix_result}
		\begin{adjustbox}{center} 
		\scalebox{1}{
		\begin{threeparttable}
			\begin{tabular}{|c|ccc|}
				\hline
				\multirow{2}{*}{\textbf{Forked Project}} & \multicolumn{3}{c|}{\textbf{\# Fixed Cases}}                                                    \\ \cline{2-4} 
				                                         & \textbf{Detected}      & \textbf{Truth}         & \textbf{Err}*  \\ \hline
				Dogecoin     & 1  & 1  & -     \\
				Bitcoin Cash & 23 & 25 & (2;-) \\
				Litecoin     & 22 & 22 & -     \\
				Bitcoin SV   & 1  & 1  & -     \\
				Dash         & 11 & 10 & (-;1) \\
				Zcash        & 2  & 1  & (-;1) \\
				Bitcoin Gold & 14 & 14 & -     \\
				Horizen      & 1  & -  & (-;1) \\
				Qtum         & 28 & 28 & (1;1) \\
				DigiByte     & 14 & 14 & -     \\
				Ravencoin    & 3  & 3  & -     \\ \hline
				\multirow{2}{*}{\textbf{Sum}}            & \multirow{2}{*}{\textbf{120}} & \multirow{2}{*}{\textbf{119}} & \multirow{2}{*}{\textbf{(3;4)}} \\
				             &    &    &       \\ \hline  \hline
				Binance      & 5  & 5  & -     \\
				Avalanche    & 3  & 3  & -     \\
				Polygon      & 6  & 6  & -     \\
				Celo         & 4  & 4  & -     \\
				Optimism     & 1  & 1  & -     \\ \hline
				\multirow{2}{*}{\textbf{Sum}}            & \multirow{2}{*}{\textbf{19}}  & \multirow{2}{*}{\textbf{19}}  & \multirow{2}{*}{\textbf{-}}     \\
				             &    &    &       \\ \hline
			\end{tabular}
		\begin{tablenotes}
	    \item \small * represents (the number of missed cases; the number of mistake cases).
	    \end{tablenotes}
		\end{threeparttable}
		}
		\end{adjustbox}
	\end{subtable}
\end{table*}

%% file: setup.tex
\subsection{Experimental Setup}
\label{sec:setup}


To make sure that \name's vulnerability detection results are reliable, we not only run \name in our experiment but also compare it with the open-source state-of-the-art ReDeBug~\cite{Redebug2012} using the same dataset and environment below.
Note that we also considered other clone detection tools (e.g.,~\cite{MVP2020, VUDDY2017, CCFinder02, Sourcercc2016}) for more comparison but eventually did not choose them for two reasons.
First, MVP~\cite{MVP2020} was not open-source and it does not support the Go language.
While VUDDY~\cite{VUDDY2017} released its signature generating scripts, its most important vulnerability search engine was not available.
Indeed, we contacted the VUDDY team and confirmed that their cloud version currently supports only one CVE in our dataset.
Second, CCFinder~\cite{CCFinder02} and SourcererCC~\cite{Sourcercc2016} are pure code clone detection tools and are not able to perform patch-based detection in our problem without adjustment.

\textbf{Dataset.}
As illustrated in \myfig~\ref{fig:workflow}, \name requires two sets of input, the target blockchain code repositories and the security patches of a reference blockchain (i.e., Bitcoin and \eth in this paper).
As a result, we collect these two sets of data as our dataset.
Specifically, for code repositories, we select all the 11 forked projects of Bitcoin from the top 100 cryptocurrencies (based on the market capitalization on CoinMarketCap) and five popular forked projects of \eth (picked from Blockscan) as our target blockchains, as previously introduced in \mysec\ref{sec:backg}. 
The total market capitalization of these 16 blockchains was around 142 billion USD.
To build a reproducible dataset, we kept a local copy of the latest version of code repositories at the time of our research on 7 September 2021 and 6 June 2022 for Bitcoin forks and \eth forks, respectively.
On the other hand, for security patches, an intuitive idea is to use the CVE (Common Vulnerabilities and Exposures) information; however, we found that there are only 12 CVEs about Bitcoin with explicit patch code and eight of them are out of the recent five years.
That said, we could select only four to test if we just use the public CVE information.

%
%

To address this problem, we select bug issues/pull requests with notable security impacts (i.e., vulnerabilities) and their patch commits (i.e., patches) directly from Bitcoin's GitHub repository according to three simple principles:
(i) the patches should be released within the recent five years since outdated patches had been applied to Bitcoin before it gets forked;
(ii) the patches that cover different vulnerability types should have a higher chance to be picked up so that we can evaluate the generality of \name;
and (iii) the patches should be applicable to most forked projects, i.e., not specific to one particular Bitcoin component or one fork.
As a result, we are able to select 32 patches of Bitcoin from June 2017 to March 2020, including four CVEs. 
For \eth, since its forks are relatively new, we select six CVEs of \eth since November 2020 as the patches.
These 38 patches involve multiple vulnerability types, including denial-of-service, race conditions, privacy leakage, and etc.
While the number of Bitcoin and \eth vulnerabilities here is not large, we have to be \textit{selective} to make sure they are actually vulnerabilities.
Indeed, Bitcoin and \eth have a limited number of vulnerabilities over the years. For example, the VUDDY dataset included only 9 CVEs of Bitcoin, with 8 of them already before 2013 and only one after 2018. Moreover, we have 16 popular forked projects of Bitcoin and \eth forked projects to test, which multiplied the total test cases to 382 ($32\times11+6\times5$).

\textbf{Environment and tool configuration.}
We evaluate \name and ReDeBug on the same virtual machine running Ubuntu 18.04 with 4GB memory configured, while the host machine is a Macbook Pro with a 3.5GHz dual-core Intel Core i7 CPU and 16GB memory.
Note that ReDeBug needs to set a \texttt{n-gram} parameter to adjust the number of lines for context code.
While the default is four, we tried from one to ten and found that when \texttt{n-gram=3}, ReDeBug achieves its best result when analyzing our dataset.

%% file: accuracy.tex
\subsection{Accuracy and Performance}
\label{sec:accuracy}


After running \name and ReDeBug on the dataset in \mysec\ref{sec:setup} (i.e., using 32 Bitcoin patches and six \eth patches to test the 16 forked projects) and performing a thorough code review of all the raw detection results (including the cases that have no any output), we are able to precisely obtain the accuracy and performance data for both tools.
Overall, \name detects 101 true vulnerabilities in 13 forked projects (Qtum, Avalanche, and Polygon do not contain any vulnerability in our dataset as we manually checked), whereas ReDeBug detects only 57 vulnerabilities in ten forked projects, which makes \name's recall 1.8 times higher than that in ReDeBug.
For performance, \name is also 1.7 times faster than ReDeBug in analyzing Bitcoin's forked projects and even 15.4 times faster in analyzing \eth's forked projects with more code per project.

\input{resources/tab/tab_vuln_type}

\mytab~\ref{tab:exp_compare} shows a breakdown of the detailed accuracy and performance results of \name and ReDeBug, where TP, FN, TN, and FP represent true positive, false negative, true negative, and false positive, respectively.
According to this table, we can calculate the precision via $TP / (TP + FP)$ and the recall via $TP / (TP + FN)$, respectively.
We find that \name achieves good precision and high recall both at 91.8\%.
In contrast, while ReDeBug has only three false positives in our dataset (mainly because it uses the exact match per code line), its recall is as low as 51.8\%.
That said, ReDeBug fails to detect many of the vulnerabilities covered by \name.
Since we aim to perform a thorough investigation of forked blockchains' vulnerabilities, \name achieves the high recall we need while introducing a low false alarming rate at only 8.18\%.
Moreover, among the 13 forked projects with vulnerabilities (i.e., no Qtum, Avalanche, and Polygon), \name detects vulnerabilities in all of them, while ReDeBug fully misses the results for two projects, Bitcoin Cash and Binance. 
In particular, \name successfully detects all the vulnerabilities in Dogecoin, Bitcoin Cash, Litecoin, Binance, Celo, and Optimism with zero false negative.

We further explore the reasons that cause \name to have a much better detection effectiveness than ReDeBug by analyzing the detailed results of detecting different clone types.
This is because while ReDeBug claims that it can handle Type-1 and Type-3 clones, the accuracy of each clone type may vary.
As shown in \mytab~\ref{tab:vuln_type}, among the 110 (TP + FN) vulnerabilities in the forked projects of our dataset, 95.5\% of them are the Type-1 and Type-3 clones, with the number of Type-3 clones slightly higher than that of Type-1 clones.
For these cases, ReDeBug achieves an accuracy of 85.7\% for Type-1 clones, but its detection rate for Type-2 and Type-3 clones drops to 0\% and 26.8\%, respectively.
This explains why ReDeBug performs better on six particular projects --- the number of Type-1 clones in those six projects (i.e., Litecoin, Dash, Bitcoin Gold, DigiByte, Celo, and Optimism) is larger than that of Type-3 clones.
Indeed, if a forked project has more Type-1 clones, it is more similar to the original project.
In contrast, \name does not have this limitation.
It is able to detect all the Type-1 clones, and misses only one and eight cases for the more complicated Type-2 and Type-3 clones, respectively. 
This indicates that \name still reaches a high rate of 80\% for Type-2 clones and 85.7\% for Type-3 clones.

For performance, \name performs much faster on all the projects than ReDeBug.
In particular, \name can finish the analysis of 10 forked projects within ten seconds, while ReDeBug just finishes only one project (i.e., Bitcoin SV) within ten seconds. 
We further analyze whether the project's LOC affects the performance of \name and ReDeBug.
For \name, we notice that it takes almost the same time (10.5s vs. 10.6s) to analyze Bitcoin Cash and Bitcoin SV, even though the LOC of Bitcoin Cash is 2.7 times that of Bitcoin SV (607K vs. 221.1K).
In contrast, the processing time of ReDeBug for the same two projects is 22.2s and 9.9s, respectively.
The difference of 2.2 times is close to the ratio of those two projects' LOC.
This indicates that the project's LOC does not explicitly affect the processing time of \name, while it has a significant effect on ReDeBug's performance.

Indeed, when we compare the performance of \name between Bitcoin forks (with fewer LOC) and \eth forks (with more LOC), we notice that \name can finish the analysis of \eth forks even faster.
It suggests that for \name, the number of target patches (32 for Bitcoin vs. 6 for \name) has a more noticeable impact on its performance than LOC.
ReDeBug, on the other hand, is the opposite, with LOC having much more impact than the number of target patches on its performance.
For example, for Qtum and Binance that have almost the same LOC, the analysis time of ReDeBug is also almost the same (33.5s vs. 30.2s).
As we mentioned earlier, typical code clone detection tools like ReDeBug perform a whole-project analysis -- so LOC dominates the performance, while \name leverages patch code contexts to search and locate only potentially relevant code for comparison -- so LOC has a much limited effect.


%% file: resources/tab/tab_vuln_type.tex
\begin{table}[t!]
	\centering
    \vspace{-1ex}
	\caption{\# of different vulnerability types in each project.}
	\label{tab:vuln_type}
	\begin{adjustbox}{center} 
	\scalebox{1}{
    \begin{threeparttable}
		\begin{tabular}{|c|rl|rl|rl|rl|}
            \hline
            \multirow{2}{*}{\textbf{Forked Project}} &
              \multicolumn{2}{c|}{\textbf{Type-1}} &
              \multicolumn{2}{c|}{\textbf{Type-2}} &
              \multicolumn{2}{c|}{\textbf{Type-3}} &
              \multicolumn{2}{c|}{\textbf{Sum}} \\ \cline{2-9} 
             &
              \textbf{T} &
              \textbf{B;R} &
              \textbf{T} &
              \textbf{B;R} &
              \textbf{T} &
              \textbf{B;R} &
              \textbf{T} &
              \textbf{B;R} \\ \hline
            Dogecoin     & 6 & (6;4) & - & -     & 10 & (10;3) & 16 & (16;7) \\
            Bitcoin Cash & 1 & (1;-) & - & -     & -  & -      & 1  & (1;-)  \\
            Litecoin     & 5 & (5;5) & - & -     & 1  & (1;-)  & 6  & (6;5)  \\
            Bitcoin SV   & 1 & (1;-) & - & -     & 11 & (10;2) & 12 & (11;2) \\
            Dash         & 7 & (7;7) & - & -     & 3  & (2;-)  & 10 & (9;7) \\
            Zcash        & 1 & (1;-) & 2 & (1;-) & 8  & (7;1)  & 11 & (9;1)  \\
            Bitcoin Gold & 9 & (9;8) & - & -     & 2  & (1;2)  & 11 & (10;10) \\
            Horizen      & - & -     & 2 & (2;-) & 9  & (7;1)  & 11 & (9;1)  \\
            Qtum         & - & -     & - & -     & -  & -      & -  & -      \\
            DigiByte     & 7 & (7;7) & 1 & (1;-) & 3  & (2;3)  & 11 & (10;10) \\
            Ravencoin    & 7 & (7;7) & - & -     & 8  & (7;3)  & 15 & (14;10) \\ \hline 
            \textbf{Sum} &
              \textbf{44} &
              \textbf{(44;38)} &
              \textbf{5} &
              \textbf{(4;-)} &
              \textbf{55} &
              \textbf{(47;15)} &
              \textbf{104} &
              \textbf{(95;53)} \\ \hline \hline
            Binance      & - & -     & - & - & 1 & (1;-) & 1 & (1;-) \\
            Avalanche    & - & -     & - & - & - & -     & - & -  \\
            Polygon      & - & -     & - & - & - & -     & - & -      \\
            Celo         & 1 & (1;1) & - & - & - & -     & 1 & (1;1) \\
            Optimism     & 4 & (4;3) & - & - & - & -     & 4 & (4;3) \\ \hline
            \textbf{Sum} &
              \textbf{5} &
              \textbf{(5;4)} &
              \textbf{-} &
              \textbf{-} &
              \textbf{1} &
              \textbf{(1;-)} &
              \textbf{6} &
              \textbf{(6;4)} \\ \hline
        \end{tabular}%
    \begin{tablenotes}
    \item \small T, B, and R represent: the total number of vulnerabilities of each clone type, the number of vulnerabilities detected by \name, and the number of vulnerabilities detected by ReDeBug, respectively.
    \end{tablenotes}
    \end{threeparttable}
	}
	\end{adjustbox}
  \vspace{-2ex}
\end{table}

%% file: result.tex
\subsection{Analysis of the Detected Vulnerabilities}
\label{sec:result}

Since \name detects not only the cloned vulnerabilities but also whether a patch is applied, we perform an analysis on both the detected vulnerabilities and the fixed cases in this subsection.
For a deep investigation on the individual vulnerability, we present it later in \mysec\ref{sec:investigate}.

As shown in \mytab~\ref{tab:exp_compare}, Bitcoin's forked projects have a total of 104 vulnerabilities.
Among the 11 projects, only Bitcoin Cash and Qtum have few vulnerabilities, while eight projects have at least 10 vulnerabilities each out of the 32 patches investigated.
In particular, Dogecoin and Ravencoin did not patch around half of the total 32 vulnerabilities.
On the contrary, \eth's forks present a better result, with only Optimism having four vulnerabilities out of the six patches investigated.
The other four projects have at most one vulnerability each, with Avalanche and Polygon fully patched.

For the result of fixed cases, the forked projects of Bitcoin and \eth have fixed a total of 138 vulnerabilities (119 for Bitcoin and 19 for \eth).
While Bitcoin's 11 forked projects have fixed 119 vulnerabilities, five of them, Dogecoin, Bitcoin SV, Zcash, Horizen, and Ravencoin, fixed only six vulnerabilities in total.
Three projects, Qtum, Bitcoin Cash, and Litecoin, contribute to 63\% of all the fixed cases.
Similar to the result above regarding the vulnerable cases, \eth's forked projects also perform better in the fixed cases.
While Optimism fixed only one vulnerability, the other four projects have fixed at least half of the investigated patches.
Indeed, when comparing the ratio of the fixed/vulnerable cases between Bitcoin's and \eth's forked projects --- 119/104 vs. 19/6, we notice that \eth's forks are more active in fixing propagated vulnerabilities.
Another aspect for measuring the project's activeness on patching vulnerabilities is the patch delay, which we provide a detailed analysis in \mysec\ref{subsec:patch_delay}.

%% file: report.tex
\subsection{Vulnerability Reporting and Response}
\label{sec:report}

As an ethical research and one contribution of this paper, we have spent significant efforts reporting all the 110 discovered vulnerabilities (including 101 TP automatically detected by \name and 9 FN manually identified by us during evaluation) to the developers of the affected forked projects via multiple channels.
In \mytab~\ref{tab:vuln_report}, we summarize the latest developers' response and actions to our vulnerability reports as of 26 July 2022.
Specifically, ``Fixed'' means that the vendor has adopted our reports to fix the issues,
``Accepted'' represents that the developers accepted our reports and were exploring appropriate patch migration,
``ACK'' suggests that the vendor has acknowledged our reports but did not explicitly indicate to fix the issues,
``Pending'' means that we have not received any response yet,
and lastly, ``Reject'' means that the vendor has denied and refused to fix the vulnerabilities.
We can see that around 74 of our 110 vulnerability reports received positive response, which demonstrates that the impact of our work.
We further classify developers' response into three categories:

\input{resources/tab/tab_vuln_report}

\textbf{Positive/Active Response.}
    Among the 13 forked projects with vulnerabilities, around half of them responded to our vulnerability report positively,
    namely Dogecoin, Ravencoin, Dash, Bitcoin Gold, Litecoin, and Binance.
    Specifically,
    Dogecoin acknowledged all of our reports and quickly fixed 11 serious vulnerabilities, while the others are scheduled or under the community discussion for appropriate patch migration.
    Meanwhile, Ravencoin accepted nearly all the reports. The developers fixed nine of them and acknowledged three except one rejection and one pending due to the compatibility consideration.
    Similarly, the developers of Dash approved nearly all the reports and informed us that they had fixed five vulnerabilities under the development branch, which will be merged into a new release in the future.
    Bitcoin Gold also fixed seven vulnerabilities in one release after around four months receiving our reports, with another one acknowledged and three under pending,
    while Litecoin fixed two of the vulnerabilities and claimed that they had noticed the other three. 
Lastly, Binance immediately acknowledged our report on BSC and rewarded us a bug bounty with the promise of fixing it.
During this reporting process, we found that developers are more likely to fix a vulnerability with authoritative proofs, especially those with CVE numbers.
For instance, the Dogecoin developers quickly released a new version of the Dogecoin core after they fixed CVE-2021-3401 and CVE-2019-15947.
However, for the other vulnerabilities with no CVE assigned, they just acknowledged them and kept them on the to-do list.


\textbf{Neutral Response.}
In this category, developers also accepted our reports but did not have intention to fix any of them yet.
    Specifically, Bitcoin SV's developers quickly acknowledged 8 of the 12 reports, and Zcash similarly acknowledged 9 of the 11 reports.
    However, both rejected a few (2 for Bitcoin SV and 1 for Zcash) due to incompatibility, and we have not received further updates from them.
Meanwhile, Horizen acknowledged four vulnerability reports with the other seven still under pending, and Celo acknowledged the only report.

\textbf{Negative/Inactive Response.}
Unfortunately, the response from the rest of three projects is not active and worrisome.
Specifically, Bitcoin Cash, DigiByte, and Optimism did not give response to any of our reports.
The worst case is DigiByte because it ignored 11 vulnerabilities, including some critical ones like CVE-2021-3401 and CVE-2019-15947.



%% file: resources/tab/tab_vuln_report.tex
\begin{table}[t!]
    \centering
    \vspace{-1ex}
    \caption{Developers' response to our vulnerability reports.}
    \label{tab:vuln_report}
    \scalebox{1}{
        \begin{tabular}{|c|c|c|c|c|c|c|}
        \hline
        \textbf{Forked Project} & \textbf{Fixed} & \textbf{Accepted} & \textbf{ACK} & \textbf{Pending} & \textbf{Reject} & \textbf{Sum} \\ \hline
        Dogecoin                                 & 11                              & 3                                 & 2                             & -                                 & -                                & 16                            \\
        Bitcoin Cash                             & -                               & -                                 & -                             & 1                                 & -                                & 1                             \\
        Litecoin                                 & 2                               & -                                 & 3                             & 1                                 & -                                & 6                             \\
        Bitcoin SV                               & -                               & -                                 & 8                             & 2                                 & 2                                & 12                            \\
        Dash                                     & 1                               & 5                          & 3                              & 1                                 & -                                & 10                            \\
        Zcash                                    & -                               & -                                 & 9                             & 1                                 & 1                                & 11                            \\
        Bitcoin Gold                             & 7                               & -                                 & 1                             & 3                                 & -                                & 11                            \\
        Horizen                                  & -                               & -                                 & 4                             & 7                                 & -                                & 11                            \\
        Qtum                                     & -                               & -                                 & -                             & -                                 & -                                & -                             \\
        DigiByte                                 & -                               & -                                 & -                             & 11                                & -                                & 11                            \\
        Ravencoin                                & 9                               & 1                                 & 3                             & 1                                 & 1                                & 15                            \\ \hline
        \textbf{Sum}            & \textbf{30}    & \textbf{9}       & \textbf{33}  & \textbf{28}      & \textbf{4}      & \textbf{104} \\ \hline \hline
        Binance                                  & -                               & 1                                 & -                             & -                                 & -                                & 1                             \\
        Avalanche                                & -                               & -                                 & -                             & -                                 & -                                & -                             \\
        Polygon                                  & -                               & -                                 & -                             & -                                 & -                                & -                             \\
        Celo                                     & -                               & -                                 & 1                             & -                                 & -                                & 1                             \\
        Optimism                                 & -                               & -                                 & -                             & 4                                 & -                                & 4                             \\ \hline
        \textbf{Sum}            & \textbf{-}     & \textbf{1}       & \textbf{1}   & \textbf{4}       & \textbf{-}      & \textbf{6}   \\ \hline
        \end{tabular}
    }
    \vspace{-4ex}
\end{table}

%% file: investigate.tex
\section{Investigating the Propagation and Patching Processes of Discovered Vulnerabilities}
\label{sec:investigate}

In this section, we conduct a deep investigation of the vulnerabilities discovered in \mysec\ref{sec:detect}.
Specifically, in \mysec\ref{subsec:vuln_root_cause}, we aim to understand how these vulnerabilities are propagated from Bitcoin and Ethereum to their forked projects.
Furthermore, in \mysec\ref{subsec:false_detection_analysis}, we diagnose some other propagation that caused our detection to fail (both FP and FN). 
Lastly, we perform a patch delay analysis in \mysec\ref{subsec:patch_delay} to understand the patching processes of the cases that were already fixed in forked projects before our detection.

\subsection{Revealing the Vulnerability Propagation from Bitcoin/\eth to Their Forked Projects}
\label{subsec:vuln_root_cause}

\input{resources/fig/fig_root_vuln}

To reveal how a vulnerability is propagated from Bitcoin and Ethereum to the forked projects, we manually check all the 110 vulnerabilities, including 104 from Bitcoin forks and 6 from \eth forks, respectively, and categorize them into three types, as shown in \myfig~\ref{fig:vuln_root_cause}.
To simplify the description in this section, we apply ``Bitcoin'' to represent both Bitcoin and \eth, unless explicitly specified.
The first type, as illustrated in \myfig~\ref{fig:fork_type}, refers to the vulnerabilities that were introduced when the project was initially forked from Bitcoin.
For better understanding and simplicity, we call it the \texttt{fork} type.
The second type, as depicted in \myfig~\ref{fig:fetch_type}, is similar to the first type except that it fetched and merged vulnerable commits of Bitcoin afterwards.
We call it the \texttt{fetch} type.
The third type, as shown in \myfig~\ref{fig:logical_flaw}, is an advanced version of the \texttt{fetch} type.
The major difference is that vulnerabilities of this type were infected with no explicitly vulnerable commits of Bitcoin.
Typically, they are caused by the defective program design or inappropriate functionality implementation that involves multiple code commits.
We call this type the \texttt{mixed} type.
In total, we identify 41 \texttt{fork}-type, 25 \texttt{fetch}-type, and 44 \texttt{mixed}-type vulnerabilities, respectively.
We conduct case studies about these three types as follows.

\textbf{Vulnerabilities directly forked in the beginning.}
In the \texttt{fork} type, vulnerabilities were propagated into the forked projects when they forked from Bitcoin. 
Many vulnerabilities, such as CVE-2022-29177 and CVE-2021-41173 from Ethereum, or CVE-2021-3401 from Bitcoin, are the classic cloned vulnerability cases to explore \texttt{fork}-type vulnerabilities and study their propagations.
Take CVE-2021-3401 as an example.
This vulnerability first appeared in Bitcoin, but we found that it also affected three other forked projects (Dash, Ravencoin, and Bitcoin Gold) since they were initially forked from Bitcoin.
As detailed in~\cite{CVE-2021-3401-detail} and the patch code in~\cite{CVE-2021-3401-Patch-Code}, it was caused by the misuse of the Qt-framework built-in arguments.
Specifically, Bitcoin and its forked projects leverage the Qt-framework~\cite{Qt} to design their own GUI programs.
However, Qt suffered from argument misinterpretation, in which attackers can inject dangerous built-in Qt arguments, e.g., \texttt{-platformpluginpath}, into a normal Qt command to load and execute their malicious plugin code remotely.

\textbf{Vulnerabilities fetched from vulnerable commits.}
In the \texttt{fetch} type, vulnerabilities were introduced when forked projects fetch commits from Bitcoin to update their functionalities without verifying whether a commit is vulnerable or neglecting a patch from Bitcoin afterwards.
Dogecoin and DigiByte (forked from Bitcoin) were also affected by the aforementioned CVE-2021-3401 yet in this way, and Optimism (forked from Ethereum) were similarly affected by the CVEs including CVE-2020-26265, CVE-2020-26264, and CVE-2020-26260.
Taking Dogecoin as example, it fetched the vulnerable commit \texttt{202d853b}~\cite{CVE-2021-3401-Vulnerable-Commit-Code} of CVE-2021-3401 from Bitcoin that sets inappropriate arguments in the class \texttt{BitcoinApplication}, but failed to pose any security check, causing a typical \texttt{fetch} vulnerability.
This is different from the \texttt{fork} vulnerabilities because Dogecoin fetched the vulnerable code \textit{actively} instead of \textit{passively} including it.
Unfortunately, there are no specification for the developers of forked projects to use the upstream code so that it is easy to skip the security patches and fetch a vulnerable commit only.

\textbf{Vulnerabilities infected with no explicitly vulnerable commits.}
Different from the \texttt{fork} and \texttt{fetch} vulnerabilities, it is hard to locate the specific vulnerable commits that introduced vulnerabilities in the \texttt{mixed} type.
It usually contains a few consecutive or discrete commits instead of the specific one(s).
Only when \textit{all} the buggy commits were included together, a vulnerability would then appear.
Typically, in the \texttt{mixed} type, the program would still run correctly at the code level, but attackers can exploit the logical flaws.
For instance, Bitcoin PR\#16512~\cite{Bitcoin-PR-16512-Patch-Code} fixed a logical flaw where the \texttt{joinpsbts} function did not shuffle its inputs and outputs, causing a privacy leak that attackers could easily identify which outputs belong to which inputs.
This vulnerability was originated from the defect of the \texttt{joinpsbts} function implementation, instead of a certain commit that made the function vulnerable.


\subsection{Diagnosing Some Other Propagation that Caused Our Detection to Fail}
\label{subsec:false_detection_analysis}

During our investigation of vulnerability propagation, we also identified some other propagation that evaded \name's detection (FN) or caused false positives (FP).
We carefully analyze all the 18 failed detection cases (9 FPs and 9 FNs) that are listed in Table~\ref{tab:false_detection_res}, and summarize them into three types, FP-\uppercase\expandafter{\romannumeral1}, FP-\uppercase\expandafter{\romannumeral2}, and FN, as shown in \myfig~\ref{fig:vuln_flase_cause}.

\input{resources/fig/fig_root_false}

\textbf{FP-\uppercase\expandafter{\romannumeral1}: no clone, and thus no vulnerability.}
As shown in \myfig~\ref{fig:no_clone_no_vuln}, the forked project sometimes keeps its outdated code and does not clone the vulnerable commit.
As a result, it has no need to fetch the patch commit either.
However, for certain vulnerabilities, there may have multiple ways to write a security patch --- some fix the root cause while others close the attack surface.
Since \name detects the vulnerable clone based on the similarity measurement with one specific patch, it is possible that it gives false alarming if the vulnerable code could be avoided in other ways.

One notable example is CVE-2018-17145~\cite{CVE-2018-17145-detail}, which caused \name to generate four same false positives, as shown in Table~\ref{tab:false_detection_res}.
We conducted a deep analysis of this DoS vulnerability. 
We found that the root cause is a susceptible variable \texttt{m\_callbacks\_pending}, 
which was introduced in Bitcoin at the commit \texttt{08096bbb}~\cite{CVE-2018-17145-Vulnerable-Commit-Code} (but forked projects like Dogecoin did not fetch this vulnerable commit).
The size of this variable would grow unlimitedly and run out of all the host memory if attackers create flooding transactions to execute an interface function called \texttt{Inventory(inv.hash)}.
Unfortunately, Bitcoin patched this vulnerability only by deleting the unrestricted \texttt{Inventory} function.
Since Dogecoin did not clone both vulnerable and patch commits, \name identifies the unrestricted \texttt{Inventory} function and thus determines that the forked vulnerable is also vulnerable.
While the interface function is still there, there was no victim \texttt{m\_callbacks\_pending} variable in Dogecoin, making attackers cannot exploit the \texttt{Inventory} function.
There are a total of seven false positives like this, as shown in \mytab~\ref{tab:false_detection_res}.


\textbf{FP-\uppercase\expandafter{\romannumeral2}: patch outdated.}
An outdated patch means that the forked projects had fetched a patch commit but neglected its further update.
As shown in \myfig~\ref{fig:patch_oudated}, suppose there was a vulnerability in both Bitcoin and its forked project.
Bitcoin released two different version of the patch at the point A and B, respectively.
The first patch is for instant fixing while the latter for the patch update.
However, the forked project just accepted the first patch.
When \name applied the final patch (i.e., the second) to detect clones, it cannot not match the target code and trigger a false positive.

For example, \name generated a false positive for Bitcoin PR\#13808 when testing Bitcoin Cash. 
Bitcoin fixed this vulnerability by using the \texttt{shuffle}~\cite{Bitcoin-PR-13808-Patch-Code} function of the C++ standard library, which is the first patch.
However, Bitcoin later substituted the patch with the \texttt{Shuffle}~\cite{Bitcoin-PR-14624-Patch-Code} function created in Bitcoin PR\#14624.
It is an updated patch to fix the issue in a more appropriate way.
Since Bitcoin Cash adopted the first patch only and neglected the update, it caused a FP of \name.
More specifically, it means that Bitcoin Cash still uses the \texttt{shuffle} function of the C++ standard library.
When \name used the updated patch for detection, it failed because \name cannot match the patch from PR\#14624. 
Similarly, \name failed in another FP-\uppercase\expandafter{\romannumeral2} type vulnerability in Qtum from Bitcoin PR\#12561.



\input{resources/tab/tab_false_detection}


\textbf{FN: target code outdated.}
\name could also encounter false negatives when the target code where the patch applies to is outdated.
In the example of \myfig~\ref{fig:target_outdated}, point A indicates an underlying vulnerability in a Bitcoin function. 
This vulnerability is further inherited along with the development of Bitcoin at point B, and Bitcoin creates a patch at point C to fix the vulnerability located at point B.
A forked project suffers from the same vulnerability because it includes a copy of the vulnerable commit at point A.
However, the Bitcoin patch at point C can not be directly applied to the vulnerability at point A due to the inconsistent code, causing a FN. 
Specifically, \name uses the patch code at point C to search the potentially vulnerable code segments in a forked project.
If \name cannot identify any related code segments, it reports nothing and poses a false negative.

Taking Bitcoin PR\#15305 as an example, it specifies the behavior of Bitcoin nodes to disconnect a block correctly when the Bitcoin program hits exceptions. 
However, when \name applied the patch code of PR\#15305~\cite{Bitcoin-PR-15305-Patch-Code} to detect clones in Bitcoin SV, nothing outputted and a false negative appeared.
This is because that Bitcoin SV keeps the outdated code cloned from Bitcoin. 
Indeed, we checked the history of the outdated code in Bitcoin SV and found that it was a copy of an old version of Bitcoin code.
This outdated code in Bitcoin SV makes the Bitcoin patch cannot be directly applied.
In total, \name made nine such false negatives due to the outdated target code, as shown in \mytab~\ref{tab:false_detection_res}. 

\input{delay}

%% file: resources/fig/fig_root_vuln.tex
\begin{figure}[t!]
	\centering
			\begin{subfigure}{0.43\textwidth}
				\centering
				\includegraphics[width=\textwidth]{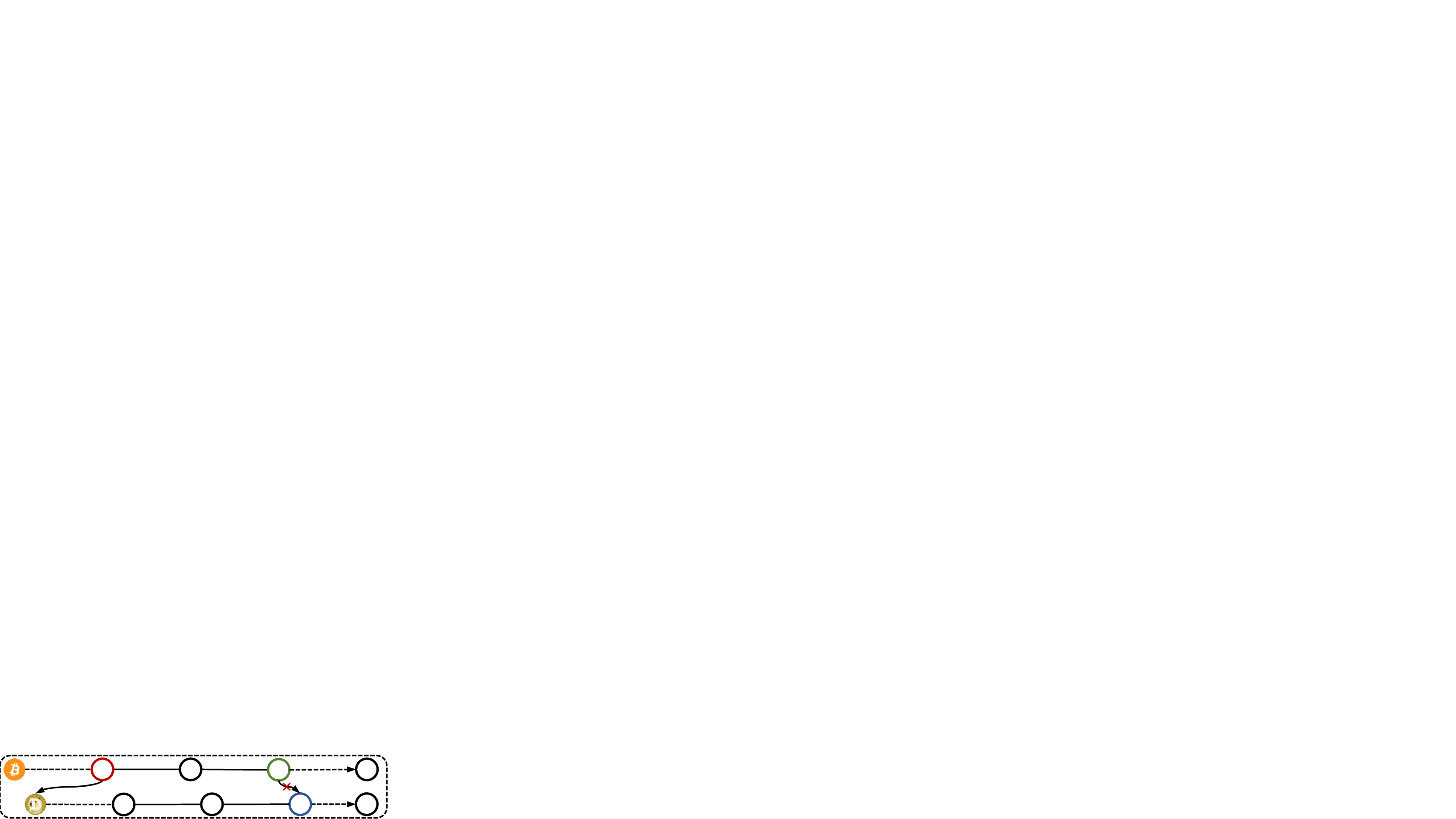}
				\vspace{-2ex}
			\end{subfigure}

			\begin{subfigure}{0.43\textwidth}
				\centering
				\includegraphics[width=\textwidth]{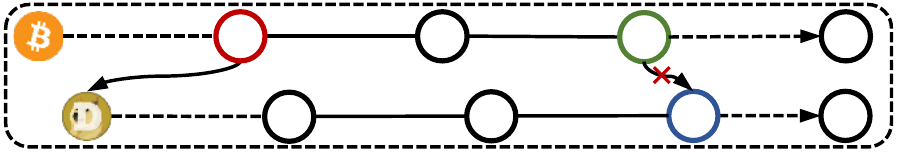}
                \caption{The \texttt{fork} type: vulnerabilities directly forked in the beginning.}
				\label{fig:fork_type}
			\end{subfigure}

			\begin{subfigure}{0.43\textwidth}
				\centering
				\includegraphics[width=\textwidth]{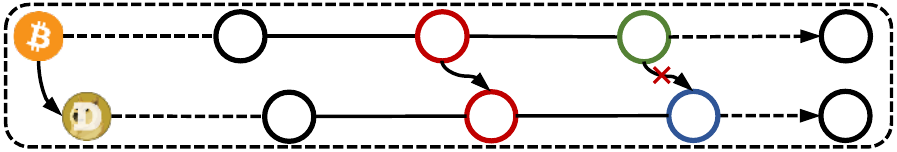}
                \caption{The \texttt{fetch} type: vulnerabilities fetched from vulnerable commits.}
				\label{fig:fetch_type}
			\end{subfigure}

			\begin{subfigure}{0.43\textwidth}
				\centering
				\includegraphics[width=\textwidth]{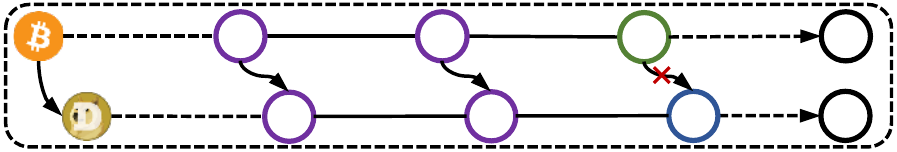}
                \caption{The \texttt{mixed} type: vulnerabilities infected with no explicitly vulnerable commits.}
				\label{fig:logical_flaw}
			\end{subfigure}

            \vspace{-2ex}
	\caption{Three types of the vulnerability propagation from Bitcoin to its forked projects.}
	\label{fig:vuln_root_cause}
	\vspace{-4ex}
\end{figure}

%% file: resources/fig/fig_root_false.tex
\begin{figure}[t!]
 	\centering
			\begin{subfigure}{0.43\textwidth}
				\centering
				\includegraphics[width=\textwidth]{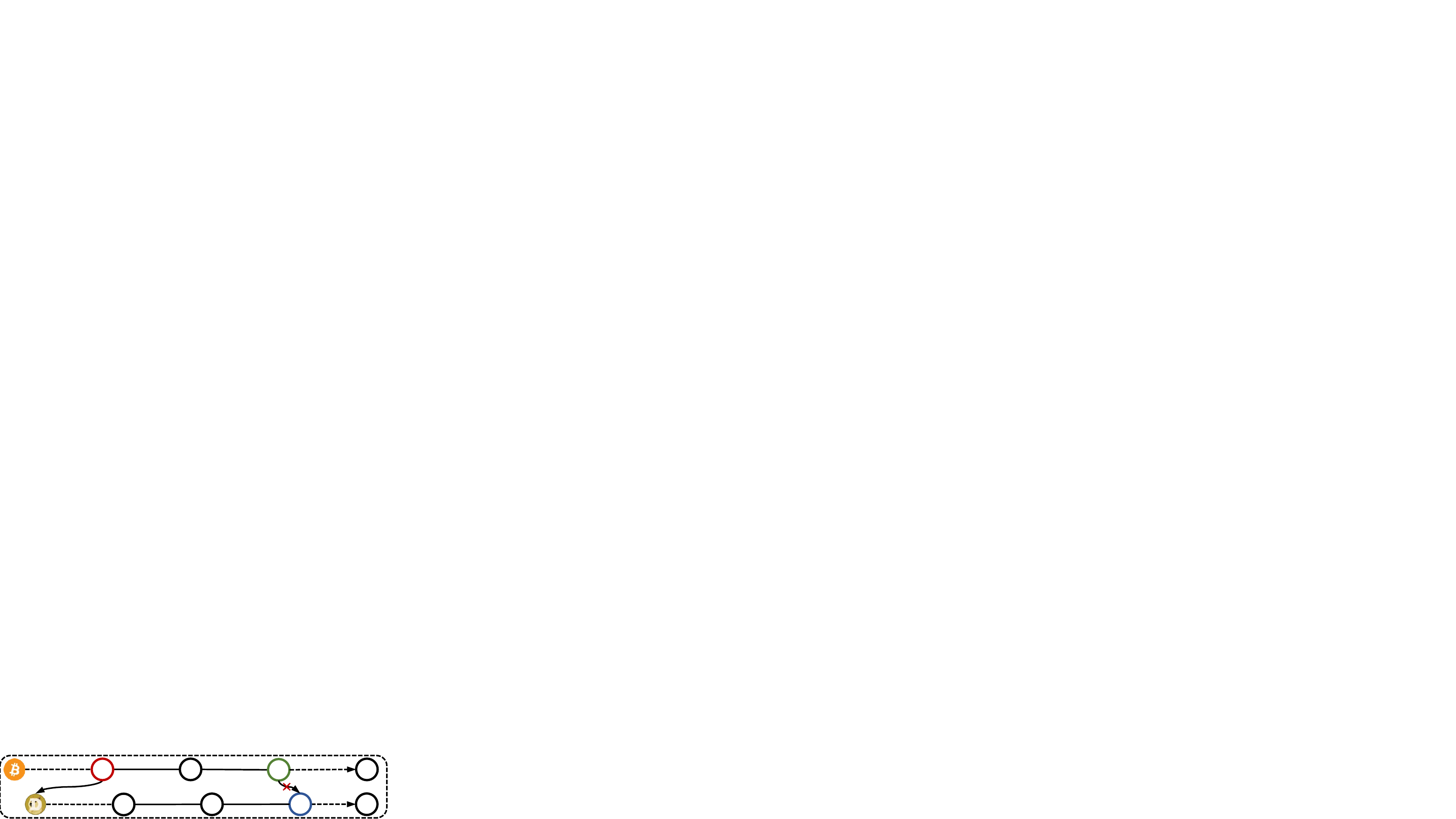}
				\vspace{-2ex}
			\end{subfigure}

			\begin{subfigure}{0.43\textwidth}
				\centering
		  		\includegraphics[width=\textwidth]{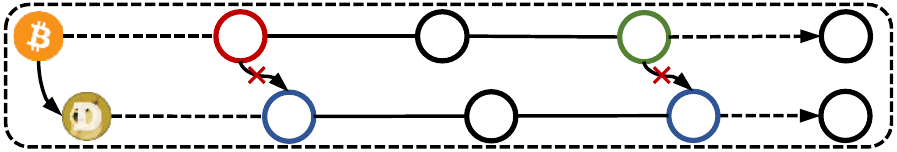}
				\caption{FP-\uppercase\expandafter{\romannumeral1}: no clone, and thus no vulnerability.}
				\label{fig:no_clone_no_vuln}
			\end{subfigure}

			\begin{subfigure}{0.43\textwidth}
				\centering
		  		\includegraphics[width=\textwidth]{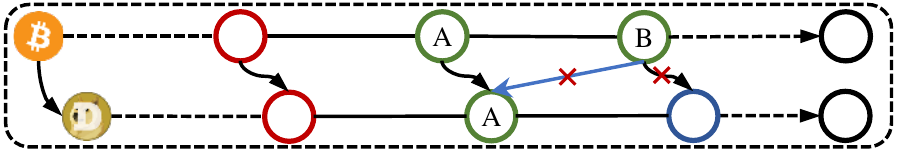}
				\caption{FP-\uppercase\expandafter{\romannumeral2}: patch outdated.}
				\label{fig:patch_oudated}
			\end{subfigure}

			\begin{subfigure}{0.43\textwidth}
				\centering
				\includegraphics[width=\textwidth]{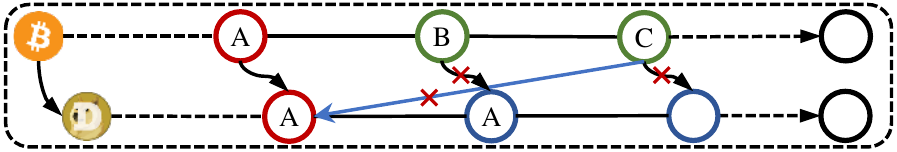}
				\caption{FN: target code outdated.}
				\label{fig:target_outdated}
			\end{subfigure}
            \vspace{-2ex}
	\caption{Three types of propagation from Bitcoin to its forked projects that caused \name to fail in terms of FP and FN.}
	\label{fig:vuln_flase_cause}
    \vspace{-4ex}
\end{figure}

%% file: resources/tab/tab_false_detection.tex
\begin{table}[t!]
    \centering
    \vspace{-2ex}
    \caption{All the 18 failed detection in \name.}
    \label{tab:false_detection_res}
    \scalebox{1}{
        \begin{tabular}{|c|c|c|c|c|}
            \hline
            \textbf{SHA} & \textbf{Source}                        & \textbf{Project} & \textbf{Cause}      & \textbf{FP/FN} \\ \hline
            beef7ec4     & CVE-2018-17145    & Bitcoin SV       & No Clone & FP-\uppercase\expandafter{\romannumeral1}            \\
            beef7ec4     & CVE-2018-17145    & Dogecoin         & No Clone & FP-\uppercase\expandafter{\romannumeral1}             \\
            beef7ec4     & CVE-2018-17145    & Horizen          & No Clone & FP-\uppercase\expandafter{\romannumeral1}             \\
            beef7ec4     & CVE-2018-17145    & Zcash            & No Clone & FP-\uppercase\expandafter{\romannumeral1}             \\
            d8318318     & CVE-2019-15947    & Bitcoin SV       & No Clone & FP-\uppercase\expandafter{\romannumeral1}             \\
            0e7c52dc     & Bitcoin PR\#12561 & Zcash      & No Clone  & FP-\uppercase\expandafter{\romannumeral1}      \\
            b8f80196     & Bitcoin PR\#14249 & Ravencoin  & No Clone  & FP-\uppercase\expandafter{\romannumeral1}      \\
            0e7c52dc     & Bitcoin PR\#12561 & Qtum             & Outdated Patch  & FP-\uppercase\expandafter{\romannumeral2}      \\
            18f690ec     & Bitcoin PR\#13808 & Bitcoin Cash     & Outdated Patch  & FP-\uppercase\expandafter{\romannumeral2}      \\
            76f74811     & Bitcoin PR\#10345 & Bitcoin SV & Outdated Target    & FN          \\
            37886d5e     & Bitcoin PR\#11568 & Horizen          & Outdated Target    & FN          \\
            37886d5e     & Bitcoin PR\#11568 & Zcash            & Outdated Target    & FN          \\
            e254ff5d     & Bitcoin PR\#13907 & Zcash      & Outdated Target    & FN          \\
            4433ed0f     & Bitcoin PR\#15305 & Horizen          & Outdated Target    & FN          \\
            effe81f7     & Bitcoin PR\#15323 & Dash       & Outdated Target    & FN         \\
            effe81f7     & Bitcoin PR\#15323 & Ravencoin        & Outdated Target    & FN          \\
            e6c58d3b     & Bitcoin PR\#15325 & Bitcoin Gold     & Outdated Target    & FN          \\
            e6c58d3b     & Bitcoin PR\#15325 & DigiByte         & Outdated Target    & FN          \\ \hline
        \end{tabular}
    }
    \vspace{-2ex}
\end{table}

%% file: delay.tex
\subsection{Patch Delay Analysis}
\label{subsec:patch_delay}

\begin{figure}[t!]
	\centering
	\begin{subfigure}{0.4\textwidth}
		\begin{adjustbox}{center}
			\includegraphics[width=\linewidth]{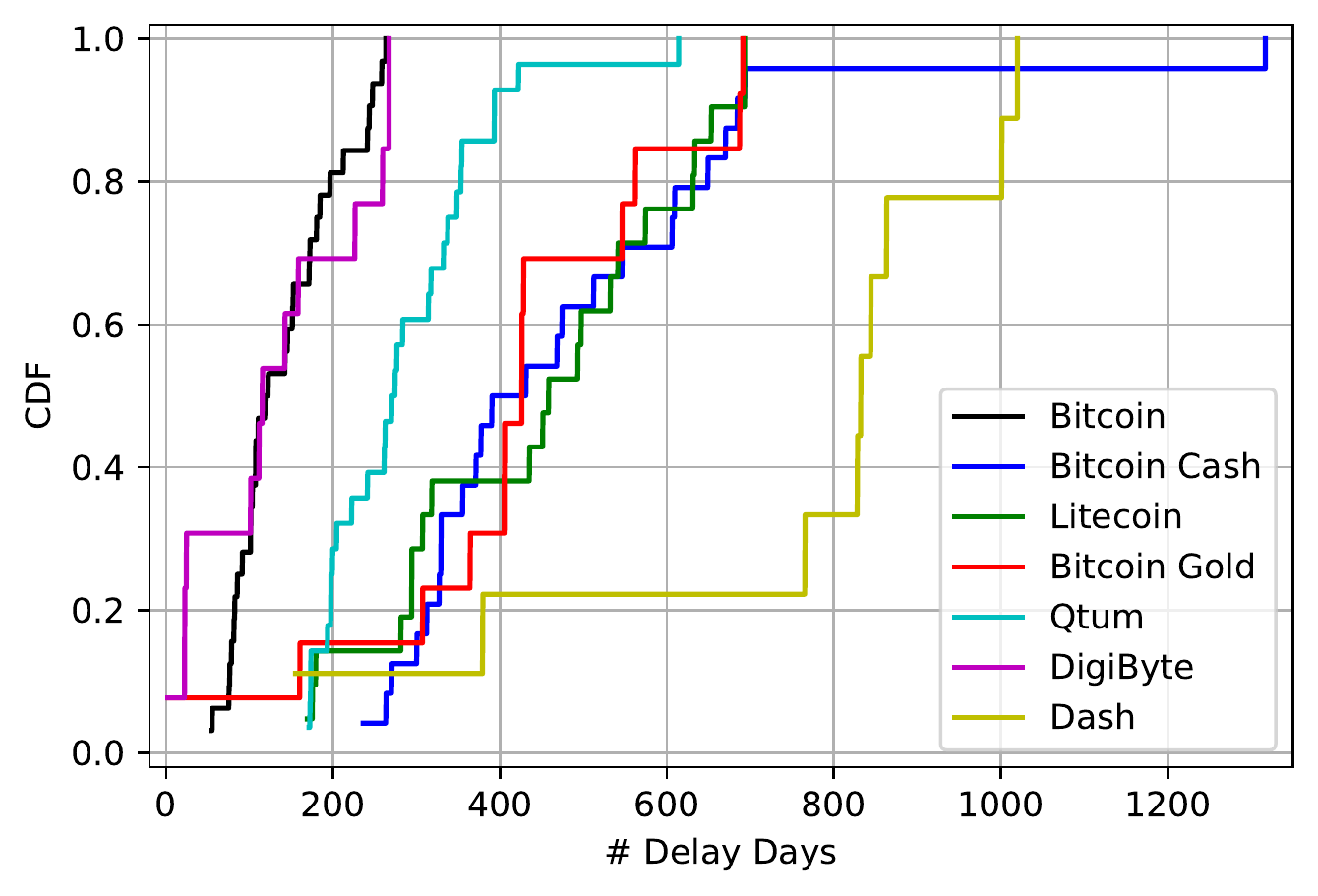}
		\end{adjustbox}
		\vspace{-3ex}
		\caption{For Bitcoin and its forked projects with enough patched cases.}
		\vspace{-1ex}
		\label{fig:repo_delay_bit}
	\end{subfigure}
	\vfill
	\begin{subfigure}{0.4\textwidth}
		\begin{adjustbox}{center}
			\includegraphics[width=\linewidth]{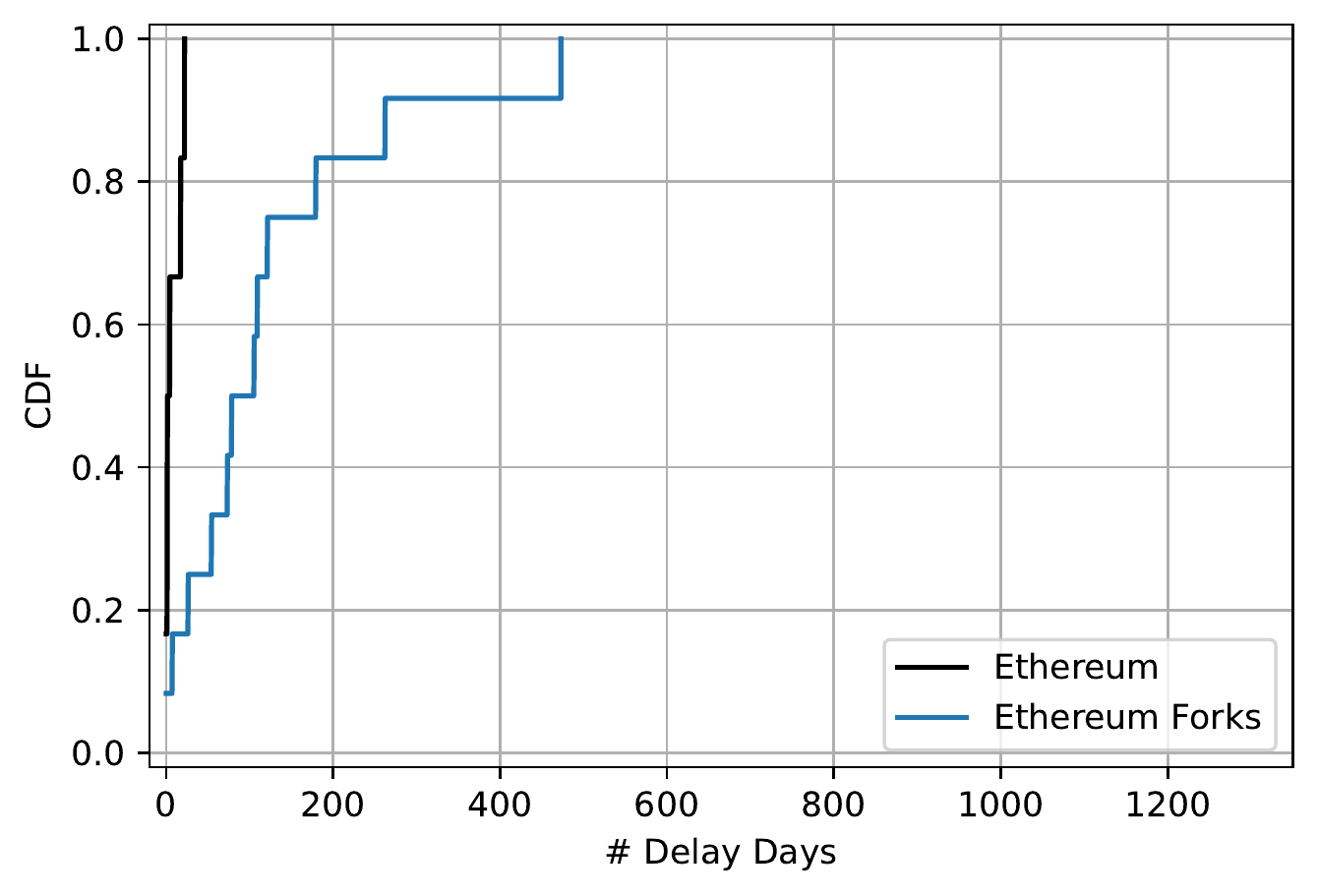}
		\end{adjustbox}
		\vspace{-3ex}
		\caption{For \eth and its forked projects as a whole.}
		\label{fig:repo_delay_eth}
	\end{subfigure}
		\vspace{-1ex}
	\caption{CDF plots of \# the delay days per security patch.}
    \vspace{-2ex}
	\label{fig:repo_delay}
\end{figure}

As previously mentioned in \mysec\ref{sec:result}, we identified a total of 138 cases (119 from Bitcoin forks and 19 from \eth forks) that were already fixed before our detection.
Among the 11 forked projects of Bitcoin, five projects have only a few fixed cases --- Dogecoin, Bitcoin SV, and Zcash have one fixed case each, and Horizen even has no fixed case.
Therefore, there is no enough data to analyze their patch delay.
Moreover, since we only investigated six patches for \eth's forked projects, i.e., they do not have many fixed cases, we put them together as ``\eth Forks''. 
Hence, we focus on the ``\eth Forks'' and six Bitcoin's forked projects with more than ten fixed cases each, i.e., Bitcoin Cash, Litecoin, Dash, Bitcoin Gold, Qtum, and DigiByte.
For each forked project, we draw a CDF plot of its patch delay days, as shown in \myfig~\ref{fig:repo_delay}.
We also plot the CDF for Bitcoin's and \eth's patch delay days, i.e., the intervals between the commit date and the release date of the patch commit in the original projects, using the black line as a reference.

According to the black line in \myfig~\ref{fig:repo_delay_bit}, Bitcoin released all the selected patches within 300 days, and 80\% of its patches were released within 200 days.
The patch delay for serious vulnerabilities is even quicker, e.g., within 110 days for the four investigated CVEs.
Unfortunately, only DigiByte can catch up with Bitcoin's release schedule, and Qtum's performance on patch delay is the second best, while the remaining projects could release only less than 20\% of the patches within 200 days.
Dash is particularly slow, with its 80\% patches released after 800 days.
In some extreme cases, the release delay could even exceed 1,000 days, e.g., in Bitcoin Cash and Dash.

The result for \eth and its forked projects is much more acceptable than Bitcoin's forked projects.
Note that we exclude Avalanche for the patch delay analysis because three of its fixed cases were included when Avalanche was first initialized.
As shown in \myfig~\ref{fig:repo_delay_eth}, for the investigated six CVEs, \eth released all the patches within a short period, at most 22 days to be specific, and four patches were released within four days.
Moreover, \eth's forked projects released all the investigated patches within 500 days, with more than 80\% released within 200 days and half of the patches released around 100 days.
Polygon is among the best, as it has six fixed cases whereas all of them were released within 110 days.

%% file: discuss.tex
\section{Discussion}
\label{sec:discuss}

In this section, we further discuss some insights and implications about the propagated vulnerabilities in forked blockchain projects, as well as their defense and detection.

\begin{figure}[h!]
	\begin{adjustbox}{center}
		\includegraphics[width=1.0\linewidth]{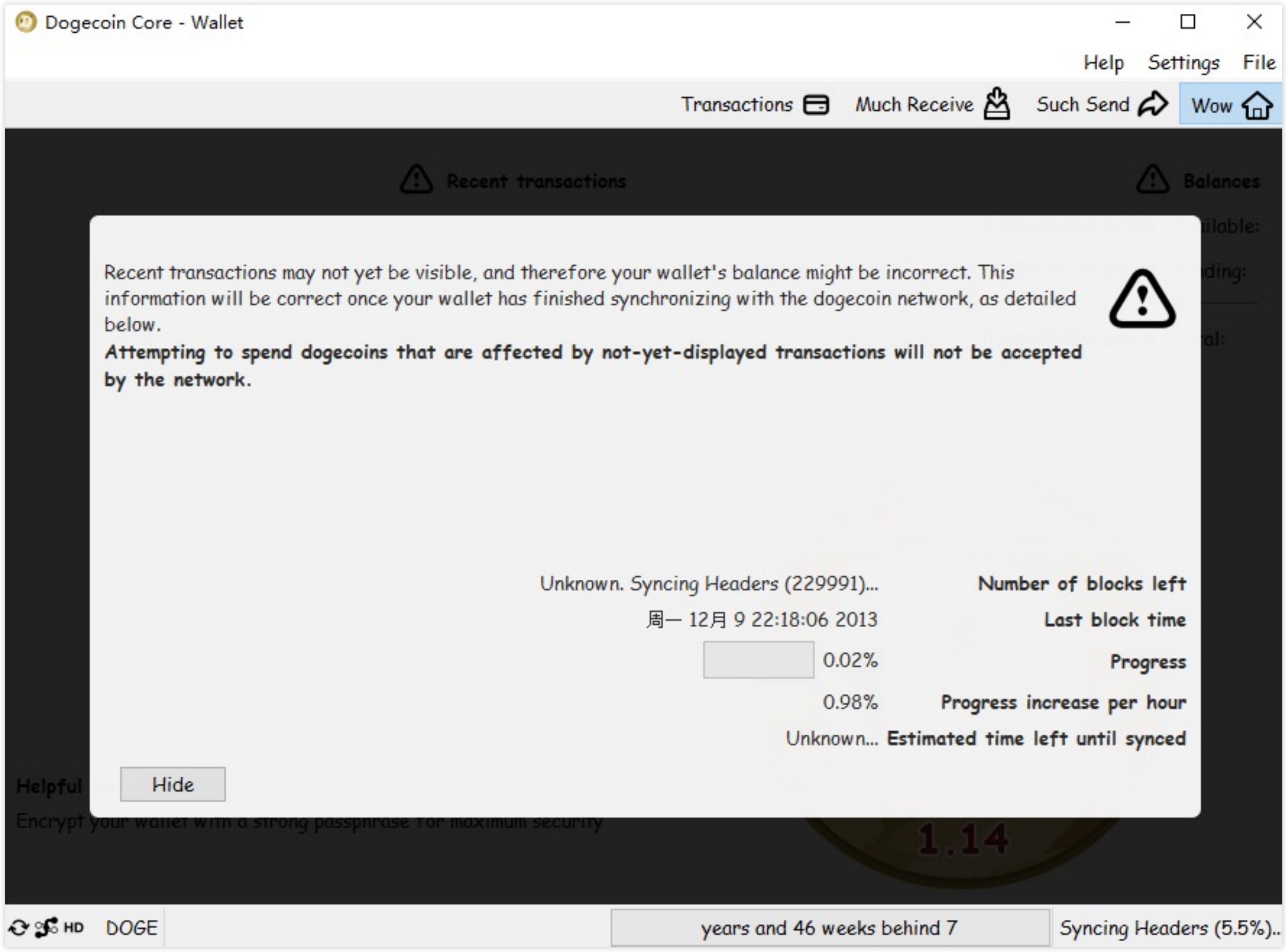}
	\end{adjustbox}
    \caption{A demo of exploiting CVE-2021-3401 in Dogecoin.} 
	\vspace{-3ex}
	\label{fig:doge_case}
\end{figure}

\textbf{Attacks against the discovered vulnerabilities.}
Since the cloned vulnerability typically has a similar code context with the original vulnerability, the vulnerability behavior is also likely to be identical.
As a result, an adversary can launch the same attack against the forked project that contains the cloned vulnerability, with only minor alterations.
For instance, in the case of CVE-2021-3401 identified in the five forked projects of Bitcoin, we followed a write-up~\cite{CVE-2021-3401-detail} that presented the details of vulnerability behavior and its complete exploitation 
and successfully exploited the discovered cloned vulnerability in all five projects.
Specifically, the root cause of CVE-2021-3401 is that the GUI program of Bitcoin (and its five forked projects) misuses Qt-framework's built-in arguments.
For example, by appending the argument \texttt{-reverse} to Dogecoin's invoking command with any wallet address, i.e., \texttt{dogecoin-qt.exe dogecoin:3E8ociqZa9mZUSwGdSmAEMAoAxBK3FNDcd -reverse}, we demonstrate that adversaries could change the program behavior by showing a reversed GUI in \myfig\ref{fig:doge_case}.
It is worth noting that real adversaries can use other more dangerous arguments, such as \texttt{-platformpluginpath} mentioned in \mysec\ref{subsec:vuln_root_cause}.
Here we use \texttt{-reverse} for easier demonstration.

Another similar case is CVE-2019-15947, which introduces the wallet information leakage problem.
Since the earlier version of Bitcoin stores \texttt{wallet.dat} unencrypted in memory, upon a crash, it may dump a core file that could be used to reconstruct users' \texttt{wallet.dat}, including private keys.
The original Bitcoin PR\#16824~\cite{Bitcoin-PR-16824} not only reported this vulnerability but also provided a shell script to exploit it.
We leveraged this script with only minor alterations to successfully exploit this vulnerability in six forked projects of Bitcoin.

\textbf{Defense or best practice for developers.}
According to our investigation in \mysec\ref{subsec:vuln_root_cause}, there are three types of propagated vulnerabilities from Bitcoin/\eth to the forked projects, i.e., the \texttt{fork}, \texttt{fetch}, and \texttt{mixed} types.
To avoid them, developers may follow the two principles: 
(i) try to avoid introducing vulnerable commits to the forked projects; 
and (ii) if a vulnerability is already introduced, developers should apply the patch code as soon as possible.
For (i), before fetching commits from the source project, developers should perform a static detection of the project (e.g., using our tool) and carefully review the commit code as well as the commit message. 
For (ii), developers should conduct security backports regularly and actively keep up with the source projects' issues/PRs.
For instance, although CVE-2021-3401 was disclosed in February 2021, the vulnerability was reported in Bitcoin PR\#16578~\cite{CVE-2021-3401-Patch-Code} much earlier on 10 August 2019.
This vulnerability would not have existed for such a long period if developers of the forked projects had followed Bitcoin's issue/PR.

\textbf{Implications on Type-4 clone detection.}
As Type-4 clones refer to semantically equivalent but syntactically different code, understanding code semantics is necessary for such detection. 
Specifically, there are two possible directions,
static/dynamic program analysis (e.g.,~\cite{ScaleClone08,RandomEquiv09,Mecc11,SliceDup01})
and machine learning on code semantics (e.g.,~\cite{SemanticLearn16,AdversClone18,ConvClone19}).
In particular,~\cite{SliceDup01} and~\cite{ScaleClone08} used the representation of isomorphic program dependence graph (PDG) as semantic clones.
Jiang et al.~\cite{RandomEquiv09}, on the other hand, regarded that given the same input, if the outputs of two code fragments are the same, they are equivalent.
Therefore, they used random testing on different code fragments and found the ones with the same behavior.
MeCC~\cite{Mecc11} proposed a memory-based approach to detect semantic clones, i.e., comparing the abstract memory states of two programs. 
Sheneamer et al.~\cite{SemanticLearn16}, CDPU~\cite{AdversClone18}, and PACE~\cite{ConvClone19} applied machine learning to detect semantic clones.
Specifically,
Sheneamer et al.~\cite{SemanticLearn16} applied classification algorithms on the extracted features of abstract syntax trees (AST) and PDG.
CDPU~\cite{AdversClone18} proposed a positive-unlabeled learning model and adversarial training to improve detection performance, 
while PACE~\cite{ConvClone19} presented an another deep learning approach by applying token-enhanced AST convolution.

%% file: related.tex
\section{Related Work}
\label{sec:related}

In this section, we review the related work on blockchain vulnerability detection and clone-based vulnerability detection.

\textbf{Blockchain vulnerability detection.}
Existing blockchain vulnerability detection mainly focused on the security of smart contracts.
For instance, Oyente~\cite{Oyente2016}, Securify~\cite{Securify2018}, ZEUS~\cite{ZEUS2018}, ETHBMC~\cite{ETHBMC20}, eThor~\cite{eThor2020}, SmarTest~\cite{SMARTEST2021}, and SAILFISH~\cite{SAILFISH2022} leveraged static analysis techniques, e.g., symbolic execution, to detect vulnerable smart contracts.
On the other hand, Sereum~\cite{Sereum2019} aimed to dynamically detect the reentrancy attacks~\cite{Reentrycy} and protect the deployed smart contracts.
Similarly,
TXSPECTOR~\cite{TXSPECTOR2020} designed a generic and flexible framework for identifying attacking transactions in Ethereum~\cite{ETHYellowPaper2022}, and SODA~\cite{SODA2020} is another generic framework for attack detection.
Lastly, Perez et al.~\cite{SmartNotExloited2021} studied the possibility of exploiting the discovered smart contract vulnerabilities.
Besides the research about smart contract vulnerability detection, DEFIER~\cite{DEFIER2021} automatically investigated the attack incidents of DApps (decentralized apps), which are built on the top of smart contracts.
Additionally, EVMPatch~\cite{EVMPatch2021} proposed a framework for instantly and automatically patching faulty smart contracts.
SolType~\cite{SolType2022} designed a refinement type system for Solidity to prevent arithmetic over- and under-flows.

However, only a few studies focused on the vulnerabilities at the system level.
Notably, Kwon et al.~\cite{BTCFAW17}, Zhang et al.~\cite{POWSEC19}, and Yang et al.~\cite{Fluffy21} investigated the consensus reward flaws and the consensus system bugs in the Bitcoin network and \eth clients, respectively.
Yi et al.~\cite{DivingBlockVuln2021} systematically mined the existing vulnerabilities from four representative blockchains, Bitcoin, \eth, Monero, and Stellar, for security insights.
Besides these works, three recent studies focused on the Bitcoin patch delay analysis that is most related to \name's \texttt{Calculator} component.
Specifically, CoinWatch~\cite{CoinWatch2020} used four CVEs of Bitcoin to test and analyze the delay of many old Bitcoin's forked projects that are no longer maintained.
It used the Simian the clone detector~\cite{Simian}, i.e., simple string match, to detect only Type-1 clones.
Similarly, Choi et al.~\cite{BTClone22} conducted a large-scale empirical analysis on the code maintenance activities of Bitcoin forks, with only limited information about security vulnerabilities.
Another technical report, GitWatch~\cite{PatchDelay22}, tried to accurately determine the patch \textit{commit} delay from Bitcoin to its forked projects.
Since \texttt{git} lacks reliable commit timestamps due to the \texttt{rebase} operation, it leveraged GitHub's event API and GitHub Archive to solve this problem.
In contrast, \name focused on the patch \textit{release} delay that does not require the \texttt{git} commit timestamp, as explained in \mysec\ref{subsec:release_delay}.


\textbf{Code clone-based vulnerability detection.} Code clone detection is an attractive research area of computer security, as it has been shown that many bugs and vulnerabilities could be cloned from one software to another~\cite{kim2018software}.
Unlike the traditional clone detection tools, such as CCFinder~\cite{CCFinder02}, CPMiner~\cite{CPMiner04}, DECKARD~\cite{deckard_2007}, and SourcererCC~\cite{Sourcercc2016}, security-oriented clone detection tools like ReDeBug~\cite{Redebug2012}, VUDDY~\cite{VUDDY2017}, MVP~\cite{MVP2020}, and VGraph~\cite{bowmanvgraph} considered both vulnerable and patched code inputs.
Specifically, ReDeBug~\cite{Redebug2012} was among the most representative works in this direction, and it has been widely used because of its generality and public code. 
Following ReDeBug, VUDDY~\cite{VUDDY2017} added variable/parameter/type/function abstraction as a preprocessing and used the generated fingerprints for more scalable code clone detection.
Similarly, MVP~\cite{MVP2020} and VGraph~\cite{bowmanvgraph} conducted more ``program analysis'' in the form of program slicing~\cite{BackDroid21} and code property graph~\cite{CodePropertyGraph14} before similarity measurement to improve the detection accuracy.
Compared with these three works, \name took a completely different path by proposing more suitable candidate code search for our problem (as in \mysec\ref{subsec:search_code}) and improving the core technique on how to better measure code similarity itself (as in \mysec\ref{subsec:code_sim}).

Recently, AI techniques have also been applied in clone-related vulnerability detection.
Specifically,
CLCDSA~\cite{CLCDSA19} utilized deep neural networks to detect cross-language code clones.
Gao et al.~\cite{SmartEmbed20} detected code clones in smart contracts by word embeddings.
Ahmadiet et al.~\cite{MLBugDetect21} leveraged machine learning-based methods to detect functionally-similar inconsistent code.
DeepBugs \cite{pradel2018deepbugs}, VulDeePecker \cite{li2018vuldeepecker}, Devign \cite{zhou2019devign}, SySeVR \cite{li2021sysevr}, and VulDeeLocator \cite{li2021vuldeelocator} utilized various kinds of code features of known vulnerabilities to train deep learning models to identify new vulnerabilities with similar code features.
Additionally, Serrano et al.~\cite{serrano2020spinfer} showed that similar yet different patches could share the same semantic and change patterns, while Zhang et al.~\cite{OEMPatchDelay21} investigated the patch delays from Android AOSP to the OEM systems. 

%% file: conclude.tex
\section{Conclusion}
\label{sec:conclude}

In this paper, we detected and investigated the vulnerabilities propagated from Bitcoin and Ethereum to their forked projects.
To this end, we proposed \name that leveraged novel context-based candidate search and a new way of calculating code similarity to efficiently and effectively identify Type-1/2/3 clones.
\name allowed us to discover 101 previously unknown vulnerabilities in 13 out of the 16 popular forked projects of Bitcoin and Ethereum, including 16 from Dogecoin, 6 from Litecoin, 1 from Binance, and 4 from Optimism.
Moreover, the evaluation showed that \name achieved good precision and high recall both at 91.8\% (1.8 times higher recall than that in the state-of-the-art ReDeBug).
We further investigated the propagation and patching processes of discovered vulnerabilities, and revealed three types of vulnerability propagation from Bitcoin/Ethereum to their forked projects, as well as the long delay (mostly over 200 days) for releasing patches in Bitcoin's forked projects (vs. $\sim$100 days for Ethereum forks).
In the future, we will continue to improve \name and expand its scope to none-blockchain domains, e.g., different Linux distributions.

%% file: appendix.tex
\appendix
\subsection{$r$'s Impact on Similarity Measurement}
\label{sec:appendix1}

\begin{figure}[b!]
	\begin{adjustbox}{center}
		\includegraphics[width=1.0\linewidth]{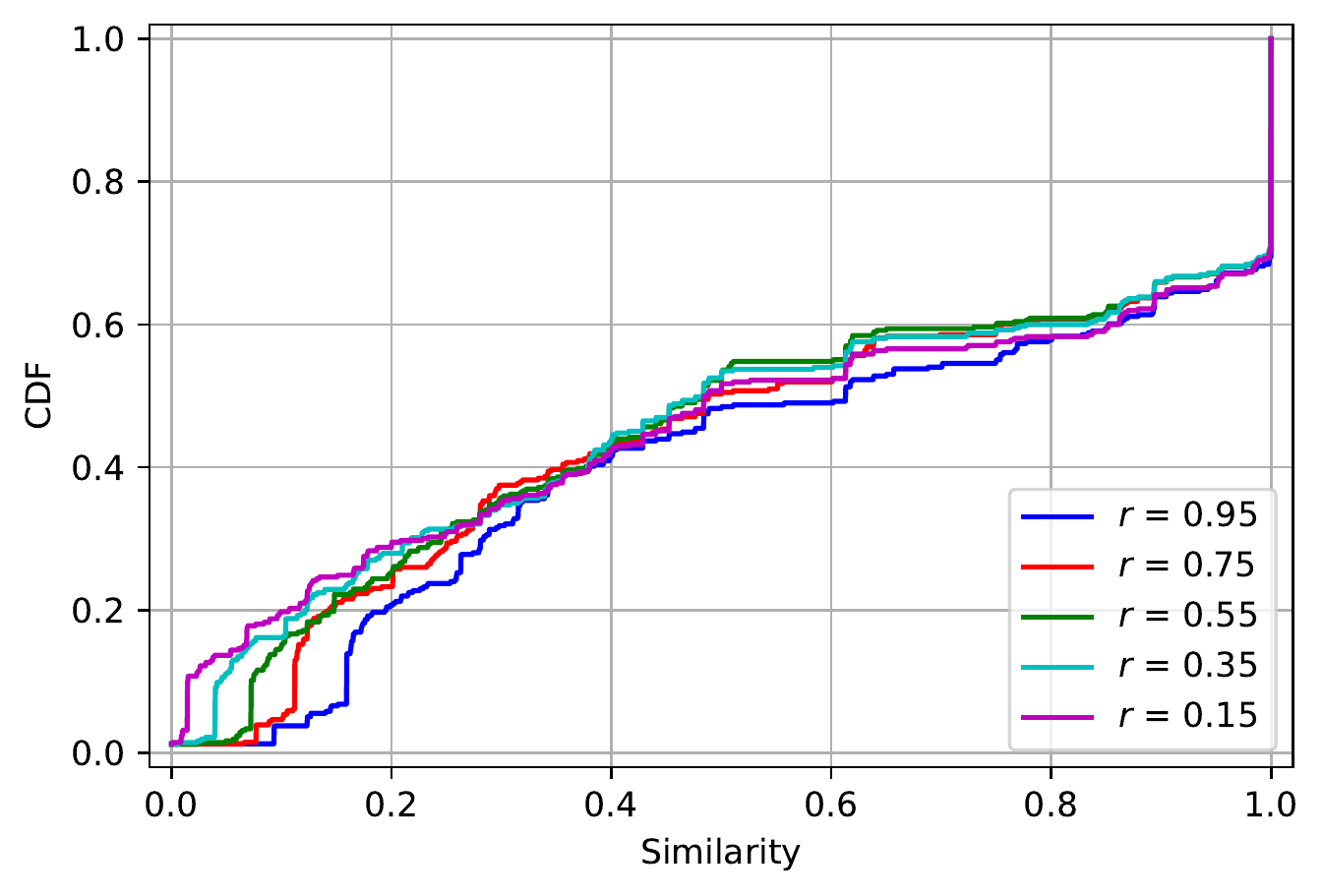}
	\end{adjustbox}
	\caption{The CDF plot of similarity with different $r$.}
	\label{fig:r_sim}
\end{figure}

As illustrated in \mysec\ref{subsec:code_sim}, we introduced the reward factor $r$ to adjust the ordering issue's influence on code similarity.
By calculating all the patch and candidate code's similarities with different $r$, we can evaluate the impact of $r$ on the similarity measurement.
In \myfig\ref{fig:r_sim}, we plot the CDF of similarity with $r$ from 0.15 to 0.95.
As we can see, $r$ has a more significant influence on the similarity when the similarity is low.
Moreover, since we try to minimize false negatives, we need to include more candidate code in the analysis.
As such, we should exclude fewer candidate code that has similarity below the threshold.
According to \myfig\ref{fig:r_sim}, when $r=0.95$, it has the least candidate code with similarity below 0.4.
Therefore, we set 0.95 as the default value of $r$.

